\documentclass[useAMS,usenatbib]{mn2e}

\usepackage{graphicx}
\usepackage{multirow} 
\usepackage{amsmath,enumerate,url} 
\def\lesssim{\la}
\def\gtrsim{\ga}

\newcommand{\appropto}{\mathrel{\vcenter{
  \offinterlineskip\halign{\hfil$##$\cr
    \propto\cr\noalign{\kern2pt}\sim\cr\noalign{\kern-2pt}}}}}

\title[]{Resonant Chains and Three-body Resonances in the Closely-Packed Inner Uranian Satellite System}
\author[Quillen \& French]{Alice C. Quillen$^1$, \& Robert S. French$^2$
\\
$^1$ Department of Physics and Astronomy, University of Rochester, Rochester, NY 14627, USA; \\
$^2$ SETI Institute, 189 Bernado Ave., Suite 100, Mountain View, CA 94043, USA \\
}

\begin{document}
\maketitle
\begin{abstract}

Numerical integrations of the closely-packed inner Uranian satellite system show 
that variations in semi-major axes
 can take place simultaneously between three or four consecutive satellites.
 We find that the three-body Laplace angle values are distributed unevenly 
 and have  histograms showing structure, 
 if the angle is associated with a resonant chain, with both pairs of bodies 
near first-order two-body resonances. 
Estimated three-body resonance libration frequencies can be only an order of magnitude lower than
those of first-order resonances. 
 Their strength arises from a small divisor from
the distance to the first-order resonances and insensitivity to eccentricity, which make up for 
their dependence on moon mass.
Three-body resonances associated with low-integer Laplace angles can also be comparatively strong
due to the many multiples of the angle contributed from Fourier components of the interaction terms. 
We attribute small coupled variations in semi-major axis,
seen throughout the simulation,
 to ubiquitous and weak three-body resonant couplings. 
 We show that a system with two pairs of
bodies in first-order mean-motion resonance can be transformed to resemble the well-studied
periodically-forced pendulum with the frequency of a Laplace angle serving as 
a perturbation frequency.
We identify trios of bodies and overlapping pairs of two-body 
resonances in each trio that have particularly short estimated Lyapunov timescales.

\end{abstract}

\section{Introduction}

Uranus has the most densely-packed system of low-mass satellites in the solar system, 
having 13 low-mass inner moons
with semi-major axes between $a = 49,752- 97,736$ km or 1.9-- 3.8 Uranian radii
\citep{smith86,karkoschka01,showalter06}.
The satellites are named after characters from Shakespeare's plays and
in order of increasing semi-major axis are Cordelia, Ophelia, Bianca, Cressida, Desdemona, Juliet, Portia,
Rosalind, Cupid, Belinda, Perdita, Puck and Mab.
External to these moons, Uranus has five larger classical moons
(Miranda, Ariel, Umbriel, Titania and Oberon)
 and a number of more distant irregular satellites.

Signatures of gravitational instability were first revealed in long-term numerical N-body integrations by \citet{duncan97}, 
who predicted collisions between Uranian satellites in only 4--100 million years. 
Observations by Voyager 2 and the Hubble
Space Telescope have shown that the orbits of the inner satellites are variable on timescales as short
as two decades \citep{showalter06,showalter08,showalter10}.
Recent numerical studies \citep{dawson10,french12} suggest that the instability is due to multiple 
mean-motion resonances between pairs of satellites.
\citet{french12} predict that the pairs Cupid/Belinda or Cressida/Desdemona
 have orbits that will cross within $10^3-10^7$ years, an astronomically short timescale.

Numerical studies of two orbiting bodies find that stable and unstable regimes
are separated by sharp boundaries (e.g., \citealt{gladman93,mudryk06,mardling08,mustill12,deck13}).
In contrast, numerical studies of closely-packed planar orbiting systems describe stability with power-law relations
 \citep{duncan97,chambers96,smith09}.
 Systems are  integrated until the orbit of one body crosses the orbit of another body
and this time,  the crossing timescale, depends on powers of the mass and 
the initial separation of the orbits \citep{duncan97,chambers96,smith09}.
The stability boundary in three-body systems is attributed to overlap of resonances involving two bodies 
\citep{wisdom80,culter05,mudryk06,faber06,mardling08,mustill12,deck13}.
In contrast, \citet{quillen11} proposed that the power law relations in multiple-body systems were due to 
resonance overlap of multiple weak three-body resonances and the strong sensitivity of these three-body resonance
strengths to masses and inter-body separations.

In this study we probe in detail one of the numerical integrations of the Uranian satellite system presented
by \citet{french12}, focusing on resonant processes responsible for instability in multiple-body systems.
In section \ref{sec:state} we describe the numerical integration and we compute
estimates for boundaries of stability.
In section \ref{sec:Ham} we construct a Hamiltonian model for the dynamics of a coplanar, low-mass
multiple-satellite or -planet system using a low-eccentricity expansion.
In section \ref{sec:twobody}
we estimate the libration frequencies of the strong two-body first-order
resonances in the Uranian satellite system.
In section \ref{sec:search3} we search for three-body resonances between bodies.
The strengths of three-body resonances that are near two-body first-order resonances
are computed in section \ref{sec:3per} and a timescale for chaotic evolution estimated for a resonant chain
consisting of pairs of bodies in mean-motion resonance in section \ref{sec:twotwo}.
In section \ref{sec:lowq} we estimate the strength of three-body resonances
that have Laplace angles with low indices.
A summary and discussion follows in section \ref{sec:sum}.

\section{The Numerical Integration and Observed Resonances}
\label{sec:state}

The numerical integration we use in this study is one of those presented and described in detail by \citet{french12}.
This simulation integrates the 13 inner moons (from Cordelia through Mab) 
in the Uranian satellite system using the SWIFT software 
 package.\footnote{SWIFT 
 is a solar system integration software package available at \url{http://www.boulder.swri.edu/~hal/swift.html.}
Our simulation uses the RMVS3 Regularized Mixed Variable Symplectic integrator \citep{levison94}.}    
The adopted planet radius is $R_U = 26,200$ km (as by \citealt{duncan97}), 
the quadrupole and octupole gravitational moments for Uranus are $J_2 = 3.34343\times10^{-3}$ 
and $J_4 = -2.885\times10^{-5}$ (as by \citealt{french91}), and the mass for Uranus is
$GM_U =5793965.663939$ km$^3$s$^{-2}$ (following \citealt{french12}).
The integrations do not include the five classical moons (Miranda, Ariel, Umbria, Titania and Oberon) 
as they do not influence the stability of the inner moons \citep{duncan97,french12}.

The masses of the inner moons
that we adopt, and specifying the integration amongst those presented by French and Showalter,
are those given in  the middle column of Table 1 of \citet{french12}.   They are estimated
from the observed moon radii assuming a density of 1.0~g~cm$^{-3}$.
Initial conditions for the numerical integration 
in the form of a state vector (position and velocity) for each moon
and dependent on the assumed moon masses, were determined through integration and  iterative orbital fitting
 and are consistent with observations for the first 24 years over which astrometry was available
 \citep{french12}.

\begin{table}
\vbox to95mm{\vfil
\caption{\large Initial integration parameters \label{tab:stuff}}
\begin{tabular}{@{}lllllll}
\hline
Satellite        & $a$(km) & $e$        &  $m$           & $n$(Hz)  & $\omega/n$ \\
\hline
Cordelia       & 49751.8 & 0.00024 & 4.47e-10 & 2.1706e-04 & 1.40e-03 \\
Ophelia        & 53763.7 & 0.01002 & 5.87e-10 & 1.9320e-04  &1.20e-03 \\
Bianca         & 59165.7 & 0.00096 & 9.50e-10 &  1.6734e-04  &9.87e-04\\
Cressida      & 61766.8 & 0.00035 & 3.33e-09 &  1.5687e-04  &9.05e-04 \\
Desdemona & 62658.3 &0.00023 & 2.07e-09 & 1.5354e-04  &8.80e-04 \\
Juliet            & 64358.3 &0.00074 & 7.18e-09 &  1.4749e-04  &8.34e-04 \\
Portia           & 66097.4 &0.00017 & 1.66e-08 &  1.4170e-04  & 7.90e-04\\
Rosalind       & 69927.0 & 0.00033 & 2.25e-09 &  1.3022e-04  &7.06e-04\\
Cupid            & 74393.1 & 0.00170 & 3.52e-11 &  1.1867e-04  &6.23e-04\\
Belinda         & 75255.8 & 0.00027& 4.40e-09 & 1.1663e-04  &6.09e-04\\
Perdita          & 76417.1 &0.00351 & 1.06e-10 & 1.1398e-04  &5.91e-04\\
Puck             & 86004.7 & 0.00009 & 2.56e-08 & 9.5457e-05  &4.66e-04 \\
Mab              & 97736.3 & 0.00246 & 8.34e-11 & 7.8792e-05  & 3.61e-04 \\
\hline
\end{tabular} 
{\\ The semi-major axis, $a$ (in km), and eccentricity, $e$, 
are initial geometrical orbital elements for the numerical integration studied here,
and presented and  described by \citet{french12}.  
The ratio of the mass of the moon to the planet  is given as $m$.
Masses are based on the observed radii assuming  a density of 1~g~cm$^{-3}$, and 
are consistent with those 
listed in the middle column of Table 1 of \citet{french12}.  
Mean motions, $n$, are in units of Hz.  The unitless
$\omega/n$ is the ratio of precession rate to mean motion.}}
\end{table}

Using the state vectors output by the integrations,
we compute the geometric orbital elements of \citet{borderies94}, 
as implemented in closed-form solution by \citet{renner06}, 
because they are not subject
 to the short-term oscillations present in the osculating elements caused by Uranus's oblateness.
 For each moon,
 initial semi-major axis, $a$, and eccentricity, $e$, are listed in Table \ref{tab:stuff}, along with 
 mean motion, $n$, secular precession frequency, $\omega$, and the ratio
 of the moon to planet mass, $m$.

\begin{figure*}
\begin{center}
\includegraphics[width=15cm]{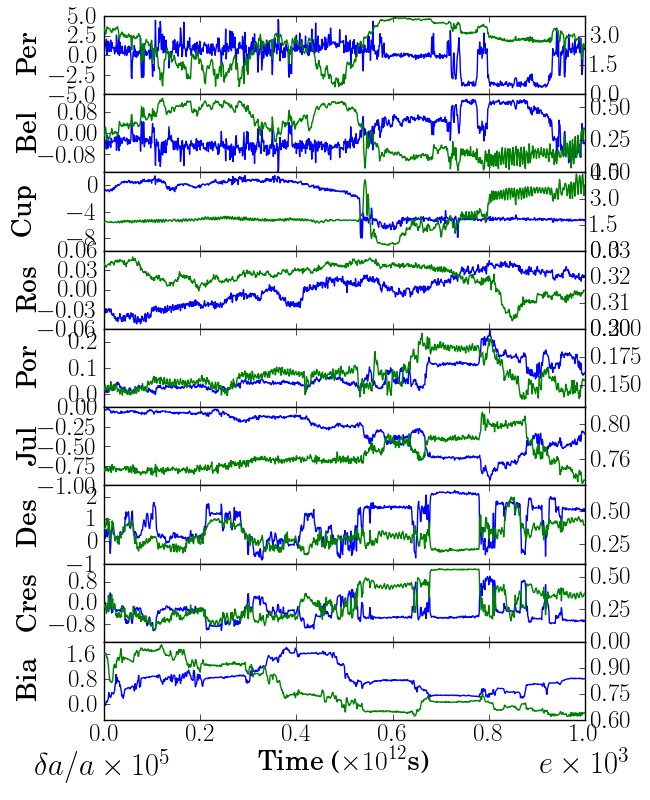}
\caption{Semi-major axes and eccentricities of the inner Uranian moons during the first part of
the numerical integration. Not all moons are plotted; see \ref{sec:state}.
Plotted as blue lines and with the $y$-axis on the left are deviations
 $(a-a_0)/a_0 \times 10^5$ where $a$ is the semi-major axis and $a_0$ is its initial value for each moon.   
The green lines show the eccentricities $\times 10^3$ with the $y$-axis on the right.
Scaling factors are written on the lower left and right. 
This figure illustrates coupled variations in semi-major axis between two, three or four bodies.
Anti-correlated variations in eccentricity and semi-major axis are evident for the lower-mass
body when two bodies are in a first- or second-order mean-motion resonance. 
}
\label{fig:geom_ae}
\end{center}
\end{figure*}

The  integration output contains state vectors for the 13 inner satellites at times separated
by $10^7$ s and the integration is $t=3.6 \times 10^{12}$s long ($1.2 \times 10^5$ yr).  
We focus  on the first part of the integration ($t<10^{12}$s), 
when the variations in the bodies have not deviated 
significantly from their initial semi-major axes and eccentricities, 
and before Cupid and Belinda enter a regime of first-order resonance overlap, jumping from resonance to resonance
(as illustrated by \citealt{french12}, see their Figures 2 and 3).
To average over short timescale variations in the orbital elements, we computed median values of 
the semi-major axes and eccentricities in time intervals
 $10^9$s long (and consisting of 100 recorded states for this integration).
These are shown to $t=10^{12}$s  in Figure \ref{fig:geom_ae}.
The semi-major axes as a function of time are plotted as a unitless ratio $(a-a_0)/a_0 \times 10^5$ 
where $a_0$ is the initial semi-major axis and the eccentricities are shown 
multiplied by $10^3$.

Figure \ref{fig:geom_ae} shows that variations in semi-major axes between bodies are correlated.
As pointed out by \citet{french12}, there are a number of strong first-order mean-motion resonances.
Cressida and Desdemona are near the 43:44 mean-motion resonance, 
Bianca and Cressida are near the 15:16 resonance, and
Belinda and Perdita are near the 43:44 resonance.
Juliet and Portia are near the 49:51 second-order mean-motion resonance.

A $p-1:p$ first-order resonance between body $i$ and body $j$ 
is described with one of the following resonant angles:
\begin{eqnarray}
\phi_{pi} &=& p \lambda_j + (1-p) \lambda_i - \varpi_i \nonumber \\ 
\phi_{pj} &=& p \lambda_j + (1-p) \lambda_i - \varpi_j \end{eqnarray}
where $p$ is an integer, $\lambda_i, \lambda_j$ are the mean longitudes of bodies $i$ and $j$.
The angles 
$\varpi_i, \varpi_j$ are the longitudes of pericenter.  
These angles move slowly when there is a commensurability between the mean motions $n_i,n_j$,
\begin{equation}
p n_j \approx (p-1) n_i.
\end{equation}
The resonant argument $\phi_{pi}$ tends to be more important when the $i$-th body is the lighter
body and $\phi_{pj}$ is more important if the $j$-th body is lighter.

In comparing semi-major axis variations with eccentricity variations, we see that
semi-major axis variations in two nearby bodies can be inversely correlated and 
the eccentricity variations of the lower-mass body
 tend to be anti-correlated with its semi-major axis variations.  
 As we will review in section \ref{sec:twobody}, within the context of a Hamiltonian model, 
when a single resonant argument  is important (that associated with $\phi_{pi}$ or $\phi_{pj}$),
conserved quantities
relate variations in the semi-major axes to the eccentricity of {\it one} of the bodies.

Figure \ref{fig:geom_ae} shows that at times
there are simultaneous variations between three or four bodies. 
The semi-major axes of 
Cressida, Desdemona, Juliet and Portia often exhibit simultaneous variations with
Cressida and Desdemona moving in opposite directions, Juliet and Portia moving in opposite
directions and Desdemona and Juliet moving in the same direction.
The correlated variations in semi-major axes seen in Figure \ref{fig:geom_ae}
between more than one body are similar to the variations exhibited
by integrated closely-spaced planetary systems (e.g., see Figure 3 of \citealt{quillen11})
that were interpreted in terms of coupling between consecutive bodies from three-body resonances. 
We will investigate this possibility below.

Eccentricities, however, are less well correlated.   Often two bodies
experience opposite or anti-correlated eccentricity variations.  For two bodies with similar masses, the two
resonant arguments, $\phi_{pi}, \phi_{pj}$, are of equal importance or strength.   Cressida and Desdemona
have similar masses and so the 46:47 resonance causes anti-correlated eccentricity variations in the two moons. 
 But rarely are eccentricity variations simultaneous among
three or more bodies.  This might be expected as the eccentricities of these satellites are low 
(see Table \ref{tab:stuff}),
and so high-order (in eccentricity) terms and secular terms
 in the expansion of the two-body interactions  in the Hamiltonian or the disturbing function are weak.

\section{Stability boundary estimates}
\label{sec:boundaries}

Here we expand on the predictions of stability estimated by \citet{french12} in their section 3.1.
A seminal stability measurement for a two-planet system is that by \citet{gladman93}.
We define a normalized distance between the semi-major axes of two bodies with semi-major axes $a_i,a_j$ as
\begin{equation}
\Delta \equiv (a_j -a_i)/a_i, 
\end{equation}
and we assume $a_i < a_j$.
Gladman's  numerical study showed that
a coplanar system with a central body and two close planets on circular
orbits is Hill stable (does not ever undergo close encounters) as long as the initial  separation 
$\Delta \lesssim \Delta_G$ with
\begin{equation}
  \Delta_G \equiv  2.4(m_i + m_j)^{1/3}.
\end{equation}
Here $m_i$ and $m_j$ are the planet masses divided by that of the central star. 

\citet{chambers96} explored equal-mass and equally-spaced
but multiple-planet planar systems finding that $\Delta \lesssim \Delta_C$ is required for Hill
stability with
\begin{equation}
\Delta_C \equiv 10 R_{mH}/a_i
\end{equation}
and $\Delta$ computed between a consecutive pair of planets.
Here the mutual Hill radius 
\begin{equation} R_{mH} \equiv \left({m_i + m_j \over 3}\right)^{1/3}  \left({a_i+a_j \over 2 }\right). \end{equation}

In the planar restricted three-body system, a low-mass object in a nearly-circular orbit near a planet
in a circular orbit is likely to experience close approaches with a planet when
$\Delta \lesssim \Delta_W$ with
\begin{equation}  
\Delta_W \equiv 1.5 m^{2/7},
\end{equation}
where $m$ is the mass ratio of the planet to the star.  This relation is known as the 2/7-th law
and the exponent is predicted by a first-order mean-motion resonance overlap criterion \citep{wisdom80}.
The coefficient predicted by \citet{wisdom80} is 1.3, but numerical studies suggest
 it could be as large as 2 \citep{chiang09}; here we have adopted an intermediate value of 1.5.
For a low-mass body apsidally aligned with a low but non-zero eccentricity planet, the 2/7-th law is unchanged 
for bodies with low initial free-eccentricity
\citep{faber06}; otherwise the chaotic zone boundary is near 
\begin{equation}
 \Delta_e \equiv 1.8 (m e)^{1/5},
\end{equation} 
where $e$ is the low-mass body's
eccentricity \citep{culter05,mustill12}.  This relation is known as the 1/5-th law.

For consecutive pairs of Uranian satellites we compute these four measures of Hill stability using initial state vectors
for each body as described in section \ref{sec:state} and listed in Table \ref{tab:stuff}.
The 2/7-th and 1/5-th laws  are derived for a massless body near a planet but here
all the bodies have mass.  
For each consecutive pair we use the maximum masses and eccentricities,
 computing the boundaries (in normalized semi-major axis) as
\begin{eqnarray}
\Delta_W &=& 1.5 \left[ \max(m_i, m_j) \right]^{2\over 7} \nonumber \\
\Delta_e &=& 1.8 \left[ \max (m_i,m_j) \max(e_i, e_j) \right]^{1 \over 5}.  \label{eqn:delta_max}
\end{eqnarray}
In section \ref{sec:twobody} below,
we estimate the first-order resonance width for two massive bodies   
and explain why we use the maximum mass in these distances.

The four measures of stability, $\Delta_G, \Delta_C, \Delta_W,$ and  $\Delta_e$ are
listed in Table \ref{tab:stab}.   We expect instability if $\Delta$ divided by any of these measures is less than 1. 
All measures of stability suggest that the inner Uranian satellite system could be stable.
However, four pairs of consecutive satellites are near estimated boundaries of instability.  These
pairs are Cressida/Desdemona, Juliet/Portia, Cupid/Belinda and Belinda/Perdita.
The stability boundaries suggest that
Cordelia and Ophelia are dynamically distant from the remaining bodies as are Puck and Mab.
Bianca through Rosalind are close together as are Cupid through Perdita. 
Cordelia, Ophelia, Puck and Mab are not plotted in Figure \ref{fig:geom_ae}  because
they exhibited minimal variations in orbital elements and lacked variations that coincided with variations
in the elements of the other moons.

An evenly spaced equal mass multiple body system with $\Delta = 1.2 \Delta_C$ and 
 a mass ratio of $10^{-9}$ has a 
 crossing timescale  $\sim 10^{10}$ orbital periods.
(from Figure 3 by \citealt{chambers96}).
Using an orbital period for Cressida of about 11 hours this corresponds to
$10^7$ years, exceeding the crossing timescales measured by
\citet{french12} by an order of magnitude (see their Table 3).
The measured crossing timescale is shorter than that of the equally spaced
system because pairs of bodies (like Cressida and Desdemona or Cupid and Belinda) are
in or near first order mean motion resonances.  They are near first order resonances
possibly because these resonances
fill a larger fraction of phase space volume when two bodies
have nearby orbits. (The measure $\Delta_W$ is related to a first order resonance overlap condition).  
The closest two bodies are usually the first to cross orbits
and so can set the numerically measured crossing timescale in a multiple body system.

\begin{table}
\vbox to 85mm{\vfil
\caption{\large Stability estimates from pairs of moons \label{tab:stab}}
\begin{tabular}{@{}llcrrrr}
\hline
\multicolumn{2}{|c|}{Pair of Moons} & $\Delta$ &${\Delta \over \Delta_G }$ 
                                                   &${\Delta \over \Delta_C }$ 
                                                    &${\Delta \over \Delta_W }$ 
                                                    &${\Delta \over \Delta_e }$  \\
 \hline
 Cordelia &   Ophelia &    0.081 &    33.2 &    11.1 &    23.3 &     7.9 \\ 
  Ophelia &    Bianca &    0.100 &    36.3 &    12.0 &    25.3 &     8.9 \\ 
   Bianca &  Cressida &    0.044 &    11.3 &     3.8 &     7.8 &     4.9 \\ 
 Cressida & Desdemona &    0.014 &     3.4 &     1.2 &     2.5 &     2.0 \\ 
Desdemona &    Juliet &    0.027 &     5.4 &     1.8 &     3.8 &     2.7 \\ 
   Juliet &    Portia &    0.027 &     3.9 &     1.3 &     3.0 &     2.3 \\ 
   Portia &  Rosalind &    0.058 &     9.1 &     3.1 &     6.5 &     5.8 \\ 
 Rosalind &     Cupid &    0.064 &    20.2 &     6.8 &    12.6 &     6.8 \\ 
    Cupid &   Belinda &    0.012 &     2.9 &     1.0 &     1.9 &     1.1 \\ 
  Belinda &   Perdita &    0.015 &     3.9 &     1.3 &     2.5 &     1.2 \\ 
  Perdita &      Puck &    0.125 &    17.7 &     5.8 &    12.3 &     7.1 \\ 
     Puck &       Mab &    0.136 &    19.3 &     6.2 &    13.4 &     8.3 \\ 
\hline
\end{tabular} 
{\\ Here $\Delta \equiv (a_{i+1} - a_i)/a_i$ gives the separation between consecutive bodies $i$ and $i+1$.
The fourth through seventh columns list $\Delta$ divided by
$\Delta_G, \Delta_C, \Delta_W$ and $\Delta_e$, delimiting different stability estimates.
None of the values listed here imply that the system will experience close encounters, though
the Cressida/Desdemona, Juliet/Portia, Cupid/Belinda, and Belinda/Perdita pairs have lowest ratios and so are pairs of moons
nearest to regions of instability.\\
}}
\end{table}


While there might be a sharp boundary between stable and unstable systems when there are
only two planets,  in a multiple-body 
system the body masses and  separations instead define an evolutionary timescale.  
With initial conditions consisting of orbits that do not intersect (when projected onto the mid-plane),
a proxy for a stability timescale is the time  
for one body to have an orbit that crosses the orbit of another body.  
This crossing timescale, measured numerically,  has been fit by a function that is proportional
to a power of the masses and a power of the interplanetary separations
\citep{chambers96,duncan97,smith09,french12}.
The numerically measured exponents in these studies are not identical and may depend
on the number of bodies in the system, initial eccentricities (e.g. \citealt{zhou07}), 
 the masses of the individual bodies when not all masses are equal,
and their initial spacings if they are not equidistant.

Chaotic diffusion occurs in regions where resonances overlap 
(e.g. \citealt{chirikov79,wisdom80,holman96,murray97,nesvorny98,murray98,quillen11,giuppone13}).   
The 2/7-th law is derived by
computing the location where first-order mean-motion resonances between two bodies in nearly circular orbits
are sufficiently wide and close together that they overlap \citep{wisdom80,deck13}.
In contrast,
\citet{gladman93} accounted for the Hill stability boundary of two-planet systems with 
an estimate for a critical value for Hill stability, 
derived by \citet{marchal82}, at which bifurcation in phase space topology occurs.

The average mass ratio of the moons from Bianca to Perdita is $\mu \approx 4 \times 10^{-9}$ so
the inner Uranian satellites are
at the low mass end of the evenly spaced equal mass  compact systems
numerically studied by \citet{chambers96}. 
Using the fitted relation by \citet{faber07} and mass ratio $\mu = 4 \times 10^{-9}$ we estimate the
crossing time for equally spaced, equal mass multiple body systems, finding  
$\sim 10^7$ orbital periods for a spacing of $\Delta = 0.014$ (similar to the closest pairs in the inner Uranian satellite system) 
and $\sim 10^{13}$ periods for $\Delta = 0.03$ (approximately the mean spacing for moons from Bianca to Perdita). 
In comparison, the crossing timescale numerically estimated by \citet{french12} is 
$\sim 10^6$ years or $\sim 10^9$ orbital periods (using an orbital period for Cressida of about 11 hours).
The closest pairs of bodies drastically lower the crossing timescale of the whole system with the closest two bodies
usually the first to cross orbits.  But integration of a close pair of bodies in isolation does not give a good
estimate for the crossing timescale in the full multiple body system.

Three-body mean motion resonances among the mean motions of an asteroid, Jupiter and Saturn are denser
than ordinary mean motion resonances \citep{nesvorny98,smirnov13}.
Overlap of three-body resonance multiplets is an important source of chaos 
in the asteroid belt \citep{nesvorny98,murray98}.
Recently, \citet{quillen11} proposed
that chaotic evolution of planar, equal mass, closely-spaced, planetary systems is due to three-body resonances 
and estimated their strengths using 
 zeroth-order (in eccentricity) two-body interaction terms.   
Crossing timescales were estimated from the time for a system to cross
into a first-order mean-motion resonance between two bodies.   The sensitivity of the three-body resonances
to interplanetary spacing and planet mass, and the associated diffusion caused by them, could account 
for the range of crossing timescales measured numerically in compact multiple-planet systems. 
Laplace coefficients 
 are exponentially sensitive to the Fourier integer coefficients and this limits
 the maximum resonance index and so the number of three-body 
 resonances that can be important in any particular system.
Equivalently, the index is truncated at smaller integers for more widely-separated bodies,
limiting the interactions between non-consecutive bodies and accounting for the insensitivity of the crossing
time scales to the number of bodies integrated \citep{quillen11}.  

Three-body resonance strengths were previously estimated by \citet{quillen11} assuming that pairs of bodies were distant
from two-body resonances.
 However, the Uranian system contains pairs 
of moons in two-body resonance and intermittent resonant behavior is clearly seen
in the numerical integrations by \citet{french12}; by intermittent behavior we mean that there are intervals of time 
with slow smooth evolution separated by intervals with rapid chaotic transitions. 
The proximity of pairs of 
bodies to the 2/7th and 1/5th law boundaries implies that even if the first
order resonances are not overlapping, the system is  strongly affected by them. 

\citet{dawson10} previously suggested that the chaotic behavior in the Uranian satellite system
 is due to this web of two-body resonances. 
To improve upon the estimate of \citet{quillen11},
we  take into account the uneven spacing and different satellite masses when estimating
three-body resonance strengths, and we also take into account the two-body mean-motion resonances.

\section{A nearly-Keplerian Hamiltonian Model for coplanar Multiple-Body dynamics}
\label{sec:Ham}

The inner moons of Uranus have low masses and eccentricities (see Table \ref{tab:stuff}), so 
a lower-order expansion in satellite mass and eccentricity should be sufficient to capture the complexity of
the dynamics. In this section we use a Hamiltonian to describe multiple-body interactions 
in such a nearly-Keplerian setting.  This approach is similar to that previously done by 
\citet{quillen11} (but also see \citealt{holman96,deck13}). For simplicity we describe our formulation
in terms of moons orbiting a central planet, but without loss of generality the same formulation could be
applied to planets orbiting a central star.

The Hamiltonian for $N$ non-interacting massive bodies orbiting a planet (and so feeling gravity only 
from the central planet) can be written as a sum of Keplerian terms
\begin{equation}
H_{Kep} =  \sum_{j=1}^N - { m_j^3 \over 2 \Lambda_j^2}
\label{eqn:H0simp}
\end{equation}
where $m_j$ is the mass of the $j$-th body divided by
the mass of planet, $M_p$.
We have ignored the motion of the planet and have put the above Hamiltonian
in units such that  $GM_p = 1$, where $G$ is the gravitational constant.
Here the Poincar\'e momentum 
\begin{equation}
\Lambda_j = m_j \sqrt{a_j},\end{equation}
where the semi-major axis of the $j$-th body is $a_j$ and
the associated mean motion is $n_j$.
This Poincar\'e coordinate is conjugate to the mean longitude, $\lambda_j$, of the $j$-th body.
The mean longitude, $\lambda_j = M_j + \varpi_j$, where $M_j$ is the mean anomaly
and $\varpi_j$ is the longitude of pericenter of the $j$-th body and we have assumed a planar
system and so neglected the longitude of the ascending node.
We also use the Poincar\'e coordinate
\begin{equation}
\Gamma_j = m_j \sqrt{a_j }(1 -\sqrt{1- e_j^2}) \approx m_j \sqrt{a_j} \frac{e_j^2}{2}, \end{equation}
where $e_j$ is the $j$-th body's eccentricity.  This coordinate
is conjugate to the angle $\gamma_j = -\varpi_j$. 
We note that the Poincar\'e momenta retain a factor of satellite's mass. 
We ignore the vertical degree of freedom.

Interactions between pairs of bodies contribute to the Hamiltonian with a term 
\begin{equation}
H_{Int} = \sum_{j>i}  W_{ij}
\end{equation}
with
\begin{equation}
W_{ij} = -{m_i m_j \over | {\bf r}_i - {\bf r}_j|  }. \label{eqn:Wij}
\end{equation}
Here ${\bf r}_i$ are the coordinates with respect to the central mass of the $i$-th body. 

We choose to work in planet-centric coordinates.  
The momenta conjugate to planet-centric coordinates
are barycentric momenta
(see for example section 4 of \citealt{duncan98} and \citealt{wisdom96} for heliocentric coordinates).  
The N-body Hamiltonian
gains an additional term $H_{drift}$,
arising from the use of the planet-centric coordinate system;
\begin{equation}
H_{drift} = {1 \over 2 M_p} \left| \sum_{i=1}^N {\bf P}_i \right|^2 \label{eqn:Hdrift}
\end{equation}
(following equation 3b of \citealt{duncan98}).
Here ${\bf P}_i$ is the barycentric momentum of the $i$-th body 
and the sum is over all bodies except the central body.  Some attention must be taken
to ensure that the above expression has units consistent with $GM_p=1$.
Expansion of $H_{drift}$ gives the indirect terms in the expansion of the disturbing function
in the Lagrangian rather than Hamiltonian setting.

The central body could be an oblate planet.  The difference between
a point mass and an oblate mass can be described with a perturbation term, $H_{ob}$, that is the sum of the
quadrupolar and higher moments of the planet's gravitational potential.
Altogether the Hamiltonian is
\begin{equation}
H  = H_{Kep} + H_{Int} + H_{Drift} + H_{ob}.
\end{equation}
An additional term could also be added to take into account post-Newtonian corrections.

\subsection{Some notation}

We focus here on the regime of closely-spaced, low-mass, planar systems.  
We define the difference of mean motions
\begin{equation}
n_{ij} \equiv n_i - n_j \sim \frac{3}{2} \delta_{ij}  \label{eqn:nij}
\end{equation}
when $\delta_{ij}$ is small.  Here
$\delta_{ij}$ is an inter-body separation with 
\begin{equation}
\delta_{ij} \equiv \alpha_{ij}^{-1} -1 \approx 1 - \alpha_{ij} \label{eqn:deltaij}
\end{equation}
 and   the ratio of semi-major axes
\begin{equation} \alpha_{ij} \equiv a_i/a_j. \end{equation}
We use a convention $a_i < a_j < a_k$ when three bodies are discussed
so that $\alpha_{ij}, \alpha_{jk} <1$.
It is convenient to define differences of longitudes of pericenter and mean longitudes
\begin{eqnarray}
\lambda_{ij} &\equiv& \lambda_i - \lambda_j  \nonumber \\
\varpi_{ij} &\equiv & \varpi_i - \varpi_j
\end{eqnarray}
for bodies $i,j$.

Interaction strengths depend on Laplace coefficients,
\begin{equation}
b_{s}^{(q)}(\alpha) \equiv {1 \over \pi} \int_0^{2\pi} {\cos (q \phi) d \phi \over (1 + \alpha^2 - 2 \alpha \cos \phi )^{s}},
\end{equation}
where $q$ is an integer and $s$ a positive half integer.
Laplace coefficients are the Fourier coefficients of twice the function
$ f(\phi) = (1 + \alpha^2 - 2 \alpha \cos \phi)^{-s}$.
As this function is locally analytic,  the Fourier coefficients decay rapidly at large $q$ and the rate
of decay is related to the width of analytical continuation in the complex plane \citep{quillen11}.
When the two objects are closely-spaced ($\alpha_{ij} \sim 1$), the Laplace coefficient can be approximated
\begin{equation} b^{(p)}_{1/2} (\alpha_{ij}) \sim 0.5 |\log \delta_{ij}| \exp (- p\delta_{ij}) \label{eqn:approx}
\end{equation}
(see equation 10 and Figure 1 of \citealt{quillen11}).


As long as the central body is much more massive than the other bodies and the bodies are not undergoing
close encounters,
the terms $H_{Int}, H_{Drift}, H_{ob}$ in the above Hamiltonian can be considered perturbations
to the Keplerian Hamiltonian, $H_{Kep}$.  
Each of these terms can be expanded in orders of eccentricity
and in a Fourier series
so that each term contains a cosine of an angle or argument, $\phi_{\bf k}$, 
that depends on a sum of the Poincar\'e angles, 
$ \phi_{\bf k} = {\bf k} \cdot (\vec \lambda, \vec \gamma)$ where ${\bf k}$ is a vector of integers
and $\vec \lambda, \vec \gamma$ are vectors of mean longitudes and negative longitudes of pericenter for all bodies.    
The coefficients for each argument are functions of the Poincar\'e momenta. 
Expansion of the pair interaction terms is referred to as
expansion of the {\it disturbing function} and is outlined in Chapter 6 of \citet{M+D} and other
texts.  The expansion is also done in their chapter 8 using a Hamiltonian approach and in terms of  
Poincar\'e coordinates for each body.
A low-eccentricity expansion for $W_{ij}$
can be put in Poincar\'e coordinates using the relations
between semi-major axis and eccentricity and Poincar\'e momenta $\vec \Gamma,\vec \Lambda$.
We focus here on low-eccentricity terms in the Hamiltonian 
or those that depend on the momenta $\Gamma_j$ to half powers
less than or equal to 2 (or eccentricity to a power less than or equal to 4).
Terms in the expansion that do not depend on mean longitudes are called {\it secular} terms.
Secular terms are divided into two classes, those that depend on longitudes of pericenter 
($\vec \varpi$) and
those that are independent of all Poincar\'e angles.  Interactions between bodies give both types of
secular perturbation terms, whereas the oblateness of the planet only affects the precession rates
and so only gives secular terms that are independent of $\vec\varpi$.

\subsection{Secular perturbations due to an oblate planet}
\label{sec:secular}

In this section we estimate
low-eccentricity secular terms in the expansion of perturbation terms
in the Hamiltonian arising from the oblateness of the planet.
Because of the low masses and eccentricities of the satellites, we neglect secular
terms arising from interactions between satellites.

A planet's oblateness causes its gravitational potential to deviate from that
of a point source, inducing quadrupolar and higher terms in the  potential.
The gravitational potential 
\begin{eqnarray}
V (r,\alpha) \approx - {1\over r}\left[1  
- J_2 \left({R_p \over r}\right)^2 P_2(\sin\alpha) \right. \\ \left.
- J_4 \left({R_p \over r}\right)^4 P_4 (\sin\alpha)\right] \nonumber 
\end{eqnarray}
where $\alpha$ is the latitude in a coordinate system aligned with the planet's rotation axis, 
 $R_p$ is the radius of the planet and we have set $GM_p =1$.
Here $J_2,J_4$ are unitless zonal harmonic coefficients and $P_n$ are Legendre polynomials of degree $n$.
Writing $r$ in terms of
 geometric orbital elements \citep{borderies94,renner06} the above expression can be expanded in 
powers of the eccentricity (after averaging over the mean anomaly),
\begin{equation}
V_o(a) +   V_{o2}(a) e^2 + V_{o4}(a) e^4.
\end{equation}
in the equatorial plane.
The potential perturbation component that is second order in eccentricity (from equation 6.255 of  \citealt{M+D}) is
\begin{eqnarray}
V_{o2}(a)& \approx&  -{1 \over 2} n^2 a^2  \left[
{3\over 2} J_2 \left({R_p \over a}\right)^2 - {9 \over 8} J_2^2 \left({R_p\over a_i} \right)^4\right. \nonumber \\
&& \left. - {15 \over 4} J_4\left({R_p \over a} \right)^4 \right]  \label{eqn:Vo2}
\end{eqnarray}
(see equations 14, 15 of \citealt{renner06} for expressions for the mean motion and other frequencies).
The $J_2^2$ term arises  from the dependence of the mean motion on the geometric orbital element $a$.

The fourth-order coefficient, $V_{o4}$, 
only depends on the $J_2$ component of the potential. 
The potential perturbation at radius $r$ 
and latitude $\alpha=0$ due to this component is
\begin{equation}
-{1\over r} {J_2\over 2} \left({R_p \over r}\right)^2.
\end{equation}
This expression is proportional to $r^{-3}$ and we expand this with a low eccentricity expansion 
(using equation 2.83 of \citealt{M+D}).  Averaging over the mean anomaly
\begin{equation}
\left({a \over r}\right)^3  \approx 1 + {3 \over 2} e^2 + {15 \over 8} e^4.
\end{equation}
The term containing $ 3e^2/2$ gives the first term in equation \ref{eqn:Vo2}, as expected.
The fourth-order term gives an additional perturbation term to the Hamiltonian that is approximately
\begin{equation}
V_{o4}(a) = -{1 \over 2} n^2 a^2 {15 \over 8} J_2  \left({R_p \over a}\right)^2 e^4. \label{eqn:Vo4}
\end{equation}

The additional terms to the gravitational potential due to the oblateness of
the planet can be incorporated as a perturbation term, $H_{ob}$, to the 
Hamiltonian.   These terms are equivalent to the potential energy perturbation
terms given above (equations \ref{eqn:Vo2} and \ref{eqn:Vo4})
times the planet mass.   To fourth order in eccentricity we gain perturbations to the Hamiltonian
\begin{equation}
H_{ob} \approx \sum_i \left( A_{ob,i}  \Gamma_i^2 + B_{ob,i} \Gamma_i\right)
\end{equation}
in terms of the Poincar\'e coordinate $\Gamma_i$,
with coefficients for each orbiting body
\begin{eqnarray}
A_{ob,i} &=& -\frac{15 }{ 4} \frac{J_2 }{ m_i a_i^2} \left(\frac{R_p }{a_i}\right)^2 
\label{eqn:ABoblate} \\
B_{ob,i} &=& -n_i  \left[ 
{3\over 2} J_2 \left({R_p \over a_i}\right)^2 - {9 \over 8} J_2^2 \left({R_p\over a_i} \right)^4 - {15 \over 4} J_4\left({R_p \over a_i} \right)^4 \right] 
\nonumber
\end{eqnarray}
where $A_{ob,i}$ comes from equation \ref{eqn:Vo4} and $B_{ob,i}$ comes from equation \ref{eqn:Vo2}.
If desired, these coefficients can be put entirely in Poincar\'e coordinates using
$a_i = \Lambda_i^2 /m_i$.
The sign for precession $\dot \varpi_i$ is correct (and positive) as the angle $\gamma_i = - \varpi_i$ is conjugate
to the momentum $\Gamma_i$.

Using equation \ref{eqn:ABoblate} for $B_{ob,i}$, it is useful to compute the difference in precession rates
for two nearby bodies
\begin{equation}
\dot \varpi_{ij} \equiv \dot \varpi_i - \dot \varpi_j  \approx B_{ob,j} - B_{ob,i}
\approx {21 \over 4} J_2 n_i \left({R_p \over a_i }\right)^2 \delta_{ij}, \label{eqn:dotvarpiapprox}
\end{equation}
which is positive for $a_j > a_i$ as the precession rate is faster for the inner body than the outer body.

\section{Two-body first-order mean-motion resonances}
\label{sec:twobody}

In this section we estimate the size scale of two-body mean-motion resonances in the Uranian satellite system.
When expanded to first order in  eccentricity the two-body interaction terms $W_{ij}$ 
have Fourier components in the gravitational potential 
\begin{eqnarray}
\sum_{q=-\infty}^\infty 
\left[ 
  	V_{ij,q}^i  \cos (q \lambda_j + (1-q)\lambda_i - \varpi_i)+  \qquad \qquad \right. \nonumber \\
	\qquad \qquad   \left. V_{ij,q}^j \cos (q \lambda_j + (1-q)\lambda_i - \varpi_j)
\right]
\label{eqn:Vsum}
\end{eqnarray}
where 
\begin{eqnarray}
V_{ij,q}^i &= - {m_i m_j \over a_j} e_i f_{27}(\alpha_{ij},q) \approx
-{m_i m_j^3 \over \Lambda_j^2} \left(2\Gamma_i \over  \Lambda_i \right)^{1\over 2} f_{27} (\alpha_{ij},q)    \nonumber \\
 V_{ij,q}^j &=  -{m_i m_j \over a_j} e_j f_{31}(\alpha_{ij},q) \approx
 -{m_i m_j^3 \over \Lambda_j^2}\left(2\Gamma_j \over  \Lambda_j \right)^{1\over 2} f_{31} (\alpha_{ij},q)
 \nonumber \\ \label{eqn:Vij}
\end{eqnarray}
and coefficients
\begin{eqnarray} 
f_{27}(\alpha,q) &\equiv& {1 \over 2}\left[ -2 q - \alpha D \right] b_{1/2}^{(q)}(\alpha)\nonumber \\  
f_{31}(\alpha,q) &\equiv& {1 \over 2} \left[ -1 + 2 q + \alpha D  \right] b_{1/2}^{(q-1)}(\alpha),
\label{eqn:f27f31}
\end{eqnarray}
where $D \equiv {d \over d\alpha}$
(equation 6.107 of \citealt{M+D}; also see Tables B.4 and B.7).

The convention in \citet{M+D} is that $q$ is the coefficient of $\lambda_j$.  Terms are grouped so
that at first order in eccentricities, there is only one $q$ for each resonant term and $q\to -q$ gives
a different resonance ($q:q-1 \to q:q+1$).

Approximations to the Laplace coefficients for closely-spaced systems 
(equation \ref{eqn:approx}; also see \citealt{quillen11}) give
\begin{eqnarray}
f_{27}(\alpha,q)
\sim -f_{31}(\alpha,q)  
 \sim  -{1 \over 4 \delta } e^{-q\delta} 
\label{eqn:fapprox}
\end{eqnarray}
for $5 \lesssim q < \delta^{-1}$.  
Here  $\delta$ is the inter-body separation with  $\delta = \alpha^{-1} -1 \approx 1 - \alpha$. 
We define arguments
\begin{eqnarray}
\phi_{qi} &\equiv& q \lambda_j + (1-q)\lambda_i - \varpi_i \nonumber \\
\phi_{qj} &\equiv&q \lambda_j + (1-q)\lambda_i - \varpi_j. \label{eqn:phiij}
\end{eqnarray}
When $\varpi_{ij}\approx \pi$ the
two resonant terms would be in phase and have the same sign.  
They effectively add and so give a stronger resonance than when $\varpi_{ij} \approx 0$.
We have neglected secular terms from interactions between bodies, 
such as one proportional to $e_i e_j \cos \varpi_{ij}$, that
could influence the separation of the two resonances and induce eccentricity oscillations
(see \citealt{malhotra89} on the secular evolution of the five classical Uranian moons).
For the inner Uranian satellites, we found that the energy in this secular term is 
1-3 orders of magnitude weaker than that of the first-order
resonant terms; the secular terms are weak because they are second order in eccentricity.

Taking a Hamiltonian that contains perturbation components corresponding to a single $q$ and
 the Keplerian Hamiltonians for two bodies, 
\begin{multline}
 H_q(\Lambda_i, \Lambda_j, \Gamma_i, \Gamma_j; \lambda_i, \lambda_j, \gamma_i, \gamma_j)  =   \\
  -{m_i^3 \over 2 \Lambda_i^2 }- {m_j^3 \over 2 \Lambda_j^2 } + B_{i} \Gamma_i + B_{j} \Gamma_j  \\
  + \epsilon_i \Gamma_i^{1\over 2} \cos \phi_{qi} + \epsilon_j \Gamma_j^{1\over 2} \cos \phi_{qj}. 
 \label{eqn:Hq}
 \end{multline}
 with coefficients dependent only on semi-major axes (or $\Lambda_i, \Lambda_j$)
 \begin{eqnarray}
\epsilon_i &=& V_{ij,q}^i \Gamma_i^{-\frac{1}{2} } =
-{m_i m_j^3 \over \Lambda_j^2} \left(2 \over  \Lambda_i \right)^{1\over 2} f_{27} (\alpha_{ij},q)     \nonumber \\
&=& -{m_i^{1/2} m_j 2^{1/2} \over a_j a_i^{1/4}} f_{27}(\alpha_{ij},q) \nonumber \\
\epsilon_j &=& V_{ij,q}^j \Gamma_j^{-\frac{1}{2} } 
= -{m_i m_j^3 \over \Lambda_j^2} \left(2 \over  \Lambda_j \right)^{1\over 2} f_{31} (\alpha_{ij},q) \nonumber \\
&=& -{m_i m_j^{1/2} 2^{1/2} \over a_j^{5/4} } f_{31}(\alpha_{ij},q). \label{eqn:epsilonij}
\end{eqnarray}
Coefficients are computed from equation \ref{eqn:Vij}.
For closely-spaced systems and using equation \ref{eqn:fapprox}
 \begin{eqnarray}
 \epsilon_i &\approx& {m_i^{1/2} m_j 2^{1/2} \over a_j a_i^{1/4}} 
 \frac{e^{-q\delta_{ij}}} { 4 \delta_{ij} }  \nonumber \nonumber \\
\epsilon_j & \approx & -{m_i m_j^{1/2} 2^{1/2} \over a_j^{5/4} } 
 \frac{e^{-q\delta_{ij}}} { 4 \delta_{ij} } 
. \label{eqn:epsilonij_approx}
 \end{eqnarray}

For the inner Uranian moons the secular precession terms are predominantly caused
by the oblateness of the planet: $B_i = B_{ob,i}$ (equation \ref{eqn:ABoblate}).  
In planetary systems secular interaction terms usually set $B_i, B_j$.

We perform a canonical transformation using a generating function that is a function
of new momenta ($K_i,K_j,J_i,J_j$) and old angles ($\lambda_i,\lambda_j, \gamma_i, \gamma_j$)
(recall that the canonical coordinate  $\gamma_i = -\varpi_i$),
\begin{multline}
F_2(K_i,K_j,J_i,J_j;\lambda_i,\lambda_j, \gamma_i, \gamma_j)
 =  \\
  K_i (q \lambda_j + (1-q)\lambda_i - \varpi_i) +  J_i \lambda_i  \\
	+  K_j (q \lambda_j + (1-q)\lambda_i - \varpi_j)  + J_j \lambda_j 
\label{eqn:f2b}
\end{multline}
giving us new momenta  and their conjugate angles
\begin{multline}
\begin{array}{llll}
J_i =  \Lambda_i - (1-q)( \Gamma_i + \Gamma_j), & & \lambda_i   \\
J_j =  \Lambda_j - q   (\Gamma_i  + \Gamma_j),    & &\lambda_j   \\
K_i =  \Gamma_i,                                 && \phi_{qi} = q \lambda_j + (1-q)\lambda_i - \varpi_i \\
K_j =  \Gamma_j,                                 && \phi_{qj}  = q \lambda_j + (1-q)\lambda_i - \varpi_j\\
\end{array} \\
\label{eqn:f2b_coords}
\end{multline}
The mean longitudes $\lambda_i, \lambda_j$  are unchanged by the transformation.
Because $K_i = \Gamma_i$ we keep $\Gamma_i$ as a momentum coordinate.
Our new Hamiltonian in terms of our new coordinates
\begin{multline}
K(\Gamma_i,\Gamma_j,J_i,J_j;\phi_{qi},\phi_{qj},\lambda_i,\lambda_j)  = \\
 - \frac{m_i^3}{2} \left[ (1-q)(\Gamma_i+ \Gamma_j) + J_i\right]^{-2} \\
 - \frac{m_j^3}{2} \left[ q(\Gamma_i+ \Gamma_j) + J_j \right]^{-2} \\
+ \epsilon_i \Gamma_i^{\frac{1}{2}} \cos \phi_{qi} + \epsilon_j \Gamma_j^{\frac{1}{2}}\cos \phi_{qj}\\
+ B_i \Gamma_i + B_j \Gamma_j . 
\end{multline}
%
We assume that $\Gamma_i,\Gamma_j$ are small and expand the first two terms in
the Hamiltonian (equation \ref{eqn:Hq}) to second order in $\Gamma_i$ and $\Gamma_j$. 
Our new Hamiltonian (in terms of our new coordinates and to second order in $\Gamma_i$ and $\Gamma_j$)
\begin{multline}
K(\Gamma_i,\Gamma_j,J_i,J_j;\phi_{qi},\phi_{qj},\lambda_i,\lambda_j) \approx  - \frac{m_i^3}{2J_i^2} - \frac{m_j^3}{2J_j^2} \\
+ \frac{A}{2} \left( \Gamma_i + \Gamma_j \right)^2
+ b_i \Gamma_i + b_j \Gamma_j  \\
+ \epsilon_i \Gamma_i^{\frac{1}{2}} \cos \phi_{qi} + \epsilon_j \Gamma_j^{\frac{1}{2}}\cos \phi_{qj}.
\end{multline}
The coefficients
\begin{eqnarray}
A &=& -{3}\left[ \frac{m_i^3}{J_i^4} (1 - q)^2 +   \frac{m_j^3}{J_j^4} q^2 \right] \nonumber \\
b_i &=& \frac{m_i^3}{J_i^3} (1-q) + \frac{m_j^3}{J_j^3} q + B_i \nonumber \\
b_j &=&\frac{m_i^3}{J_i^3} (1-q) + \frac{m_j^3}{J_j^3} q + B_j.
\end{eqnarray}

As the Hamiltonian does not depend on angles $\lambda_i,\lambda_j$
the two  momenta $J_i,J_j$ are conserved.
This implies that variations in the semi-major axis are anti-correlated (as we saw in Figure \ref{fig:geom_ae}).
If the $\phi_{qj}$ resonance is weak then we can neglect variations in $\Gamma_j$ and vice versa if
the $\phi_{qi}$ resonance is weak.   The signs in the relations for $J_i, J_j$ in equation \ref{eqn:f2b_coords}
imply that eccentricity variations are anti-correlated with semi-major axis variations of the inner body
and correlated with semi-major axis variations in the outer body.
Examination of Figure \ref{fig:geom_ae}, for example motions of Cressida and Desdemona,
illustrate that many of the correlated variations in semi-major axis and eccentricity are consistent
with perturbations from a first-order mean-motion resonance. 

As $J_i, J_j$ are conserved,
the new Hamiltonian can be considered a function of only two momenta $\Gamma_i,\Gamma_j$
and their associated angles $\phi_{qi},\phi_{qj}$.
\begin{multline}
K(\Gamma_i,\Gamma_j;\phi_{qi},\phi_{qj}) =  \\
+ \frac{A}{2} \left( \Gamma_i + \Gamma_j \right)^2
+ b_i \Gamma_i + b_j \Gamma_j  \\
+ \epsilon_i \Gamma_i^{\frac{1}{2}} \cos \phi_{qi} + \epsilon_j \Gamma_j^{\frac{1}{2}}\cos \phi_{qj}
\label{eqn:K4d}
\end{multline}

For small $\Gamma_i,\Gamma_j$ we can approximate the conserved quantity
$J_i \sim \Lambda_{i0}$ where $\Lambda_{i0}$ is a reference or initial value,
$\Lambda_{i0} = m_i \sqrt{a_{i0}}$ where $a_{i0}$ is a reference or initial value
of the semi-major axis for the $i$-th body.   We denote $n_{i0}$ the mean motion
for this semi-major axis.
Using these reference values
\begin{eqnarray}
A &=& - 3 \left[ \frac{( 1-q)^2}{m_i a_{i0}^{2}}  + \frac{q^2}{m_j a_{j0}^{2}} \right] \nonumber\\
b_i &=& n_{i0} (1-q) + n_{j0} q + B_i \nonumber \\
b_j &=& n_{i0} (1-q) + n_{j0} q + B_j .\label{eqn:AB_2}
\end{eqnarray}

The dependence of $\epsilon_i, \epsilon_j$ on satellite masses  implies that the $\phi_{qi}$ resonance
with the inner body is strong primarily when the outer  satellite mass is large and
vice-versa for $\phi_{qj}$.  This dependence is expected based on similar 
resonant arguments for asteroids in resonances with Jupiter (outer body is more massive) 
and Kuiper belt objects in resonances with Neptune (inner body is more massive).
It may be convenient to compute a ratio of resonance strengths, $\mu_\epsilon$,
\begin{equation}
\mu_\epsilon \equiv -{\epsilon_j \over \epsilon_i} = 
-\alpha_{ij}^{{1\over 4}} \left({m_i \over m_j} \right)^{1\over 2}  {f_{31} (\alpha_{ij},q) \over f_{27}(\alpha_{ij},q)}, \label{eqn:mu_e}
\end{equation}
where the sign is chosen so that $\mu_\epsilon>0$.
For closely-spaced systems ($\alpha_{ij} \to 1$)  the coefficient
$f_{27}(\alpha,q) \sim -f_{31}(\alpha,q)$ (see equation \ref{eqn:fapprox})
and the ratio of strengths of the two terms 
\begin{equation}
\mu_\epsilon \sim  \sqrt{\frac{m_i}{m_j}}. \label{eqn:epsratio}
\end{equation}

The coefficient or frequency $b_i$ determines the distance to the $\phi_{qi}$ resonance
and similarly for $b_j$ and the $\phi_{qj}$ resonance.
The time derivative of the angle $\phi_{qi} - \phi_{qj}$ is
\begin{equation}
\dot \phi_{qi} - \dot \phi_{qj}  =  - \dot \varpi_{ij} \approx b_i - b_j =  B_{i} - B_{j},
\end{equation}
The frequency $b_i-b_j$ 
 sets the distance between the  $\phi_{qi}$ and $\phi_{qj}$ resonances.
Using equation 
\ref{eqn:dotvarpiapprox} for closely-spaced bodies near an oblate planet
\begin{equation}
\dot \phi_{qi} - \dot \phi_{qj}   \sim 5.25 J_2 \left( \frac{R_p}{a_i}\right)^2 \delta_{ij} n_i 
\end{equation}
Using $J_2$ for Uranus and a semi-major axis typical 
of the inner Uranian moons we estimate 
\begin{equation}
-\varpi_{ij} = 
\dot \phi_{qi} - \dot \phi_{qj}  \sim 0.1 \delta_{ij} n_i.
\end{equation}

As discussed from dimensional analysis \citep{henrard83,quillen06} 
there are  dominant timescales  in this Hamiltonian that set 
 characteristic libration frequencies at low eccentricity
\begin{eqnarray}
\nu_i = |\epsilon_i|^{2\over 3} |A|^{1\over 3} \qquad {\rm and} \qquad
\nu_j = |\epsilon_j|^{2\over 3} |A|^{1\over 3}  \label{eqn:omij}
\end{eqnarray}
depending upon which argument is chosen (applying equation 7 of \citealt{quillen06}). 
When the two bodies are near each other, equations \ref{eqn:epsratio} and \ref{eqn:omij} imply that
\begin{equation}
{\nu_i \over \nu_j }\sim \left({m_j \over m_i}\right)^{1\over 3}.\label{eqn:nuijratio}
\end{equation}
In the high $q$ limit and when two bodies are near each other, using equation \ref{eqn:AB_2}
\begin{equation}
A \sim -\frac{3q^2}{a_i^2} \left( \frac{1}{m_i} + \frac{1}{m_j} \right).
\end{equation}
Using equation \ref{eqn:omij} for $\nu_i,\nu_j$ and equation \ref{eqn:epsilonij_approx} for $\epsilon_i, \epsilon_j$
when the bodies are near each other we estimate
\begin{equation}
\nu_i \sim  m_j^\frac{1}{3}  \left( m_i  + m_j\right)^\frac{1}{3}
q^\frac{2}{3} \delta_{ij}^{-\frac{2}{3}}
e^{-\frac{2}{3} q \delta_{ij}} 
\label{eqn:nu_ij_near}
\end{equation}
and $\nu_j$ given by multiplying by a factor of the mass ratio to the one-third power (equation \ref{eqn:nuijratio}).
The square of these libration frequencies $\nu_i, \nu_j$, approximately 
delineates the adiabatic limit for resonance capture at low eccentricity \citep{quillen06}.
An initially low eccentricity system is unlikely to capture into resonance if
drifting (in $b_i$ or $b_j$) at a rate exceeding the square of $\nu_i$ or $\nu_j$.
By summing the frequencies $\nu_i, \nu_j$ to estimate resonant width, setting this equal
to spacing between resonances, a 2/7 law 
can be derived in the setting of two eccentricity massive bodies, 
and confirming the similar derivation by \citet{deck13}.
The sum of the two frequencies  $\nu_i + \nu_j \appropto \max (m_i, m_j)^\frac{2}{3}$, supporting the use
of the maximum of the two masses in equation \ref{eqn:delta_max}. 

The maximum or critical eccentricities ensuring resonant capture in the adiabatic regime 
(and delineating the regime of low eccentricity, \citealt{borderies84}) 
can also be set dimensionally (see equation 7 of \citealt{quillen06}) with 
\begin{eqnarray}
\Gamma_{i,crit} \equiv  \left| \frac{\epsilon_i}{A}\right|^{\frac{2}{3}} \qquad {\rm and} \qquad
\Gamma_{j,crit}  \equiv  \left| \frac{\epsilon_j}{A}\right|^{\frac{2}{3}} 
\end{eqnarray}
Using the definition for the Poincar\'e coordinates $\Gamma_i,\Gamma_j$, 
these correspond to critical eccentricities values
\begin{eqnarray}
e_{i,crit} & \equiv &\left|{\epsilon_i \over A}\right|^{1 \over 3} m_i^{-{1 \over 2}} a_i^{-{1 \over 4}} \nonumber \\ 
e_{j,crit} &\equiv & \left|{\epsilon_j \over A}\right|^{1 \over 3} m_j^{-{1 \over 2}} a_j^{-{1 \over 4}}. \label{eqn:ecrit}
\end{eqnarray}
Using the conserved quantities $J_i,J_j$ and definitions for the Poincar\'e coordinates, 
semi-major axis variations have a typical size
\begin{eqnarray}
\delta_i  &=& \frac{2(q-1)}{m_i a_i^{1/2}} \left( \Gamma_{i,crit} + \Gamma_{j,crit} \right) \nonumber \\
\delta_j &=& \frac{2q}{m_j a_j^{1/2}} \left( \Gamma_{i,crit} + \Gamma_{j,crit} \right) 
\end{eqnarray}
where $\delta_i = \Delta a_i/a_i$ and similarly for $\delta_j$.
From $\nu_i, \Gamma_{i,crit}$ we can construct a characteristic energy scale 
\begin{equation}
\varepsilon_i \equiv \nu_i \Gamma_{i,crit} = | \epsilon_i| \Gamma_i^\frac{1}{2}  = |\epsilon_i|^\frac{4}{3} |A|^{-\frac{1}{3}} 
\label{eqn:varepsilon}
\end{equation}
and likewise for $\varepsilon_j$.

 The ratio of the critical eccentricities 
\begin{equation} 
e_{i,crit}/e_{j,crit} \sim  (m_j/m_i)^{1/3} 
\end{equation}
 for closely-spaced bodies.
When $e_i \gtrsim e_{i,crit}$ then the resonant width depends on the eccentricity or $\Gamma_i$ with 
\begin{eqnarray}
\nu_{ei} \sim \sqrt{|A \epsilon_i|} \Gamma_i^{1\over 4} 
\qquad {\rm and} \qquad
\nu_{ej} \sim \sqrt{|A \epsilon_j|} \Gamma_j^{1\over 4} \label{eqn:nueij}
\end{eqnarray}
respectively when $e_j \gtrsim e_{j,crit}$.
Using equations \ref{eqn:AB_2},\ref{eqn:fapprox},\ref{eqn:epsilonij} and \ref{eqn:nueij}, 
we can approximate for two nearby objects
\begin{eqnarray}
\nu_{ei} &\sim& q \sqrt{ \left( m_i +  m_j\right) {e_i \over \delta_{ij}}} \nonumber \\
\nu_{ej} &\sim& q \sqrt{ \left(m_i +  m_j\right) {e_j\over \delta_{ij} }} \label{eqn:nueij_approx}.
\end{eqnarray}

To be in the region where the $\phi_{qi}$ resonance is strong we require that
\begin{equation} 
|b_i|\lesssim \left\{
\begin{array}{l}   \nu_i \\  \nu_{ei} \end{array}  \right.
\qquad {\rm for} \qquad 
\begin{array}{l} e_i \lesssim e_{i,crit} \\ e_i \gtrsim e_{i,crit} \end{array} \label{eqn:inres}
\end{equation}
and similarly using $b_j$ for the $\phi_{qj}$ resonance.
  The dividing line depends on the critical eccentricity ensuring capture
in the adiabatic limit ($e_{i,crit}, e_{j,crit}$; as discussed from dimensional analysis by \citealt{quillen06}).

\begin{figure*}
\begin{center}
\includegraphics[width=15cm]{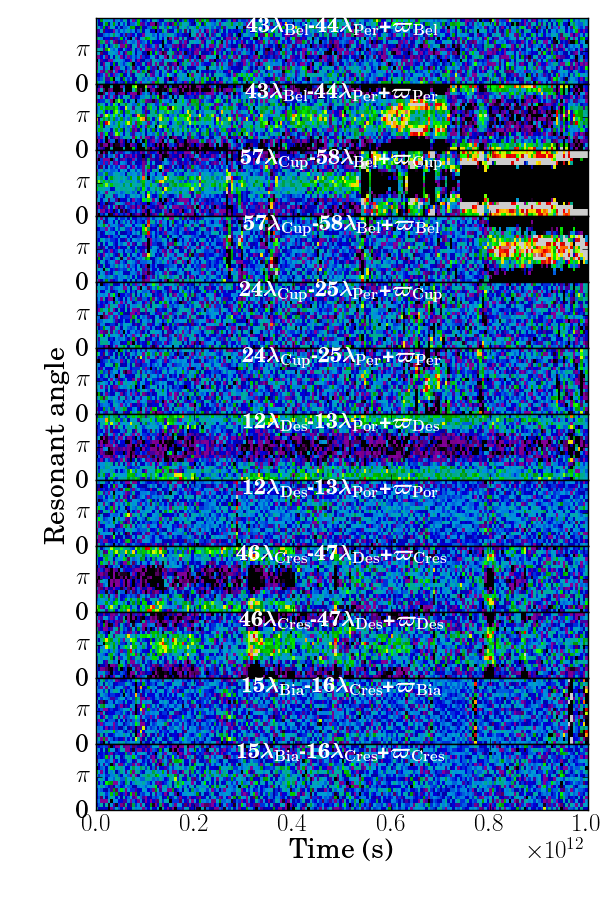}
\caption{Histograms of resonant angles associated with
first-order mean-motion resonances between two moons.
The particular resonant angle plotted is labelled in each panel.
When the color is black, the system spent no time with the resonant angle at that particular $y$ axis value.
When the color is uniformly blue, the angle was evenly distributed and the angle was circulating.
}
\label{fig:hist2body}
\end{center}
\end{figure*}

\begin{figure*}
\begin{center}
\includegraphics[width=15cm]{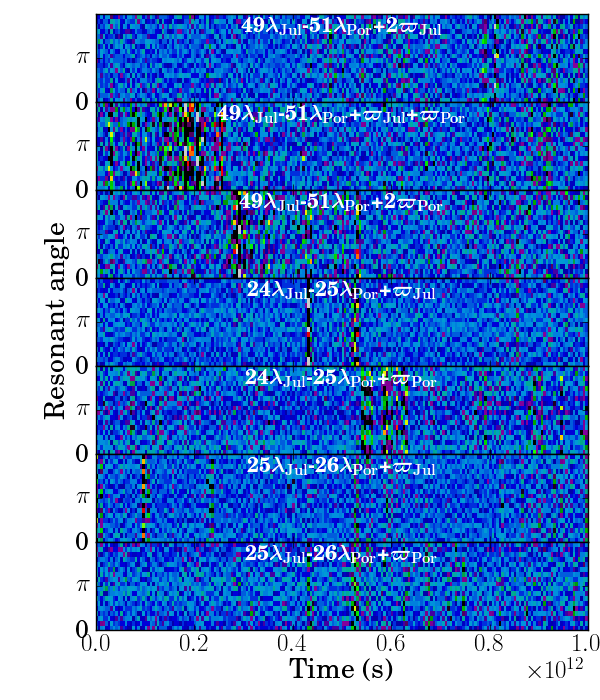}
\caption{Histograms of resonant angles associated with
first- and second-order mean-motion resonances between Juliet and Portia.
The angle $24\lambda_{Jul} - 25\lambda_{Por} + \varpi_{Por}$ spends more time at 0
and the angle $25\lambda_{Jul} - 26\lambda_{Por} + \varpi_{Por}$ spends more time at $\pi$.
Three histograms show angles associated with the 49:51 second-order mean-motion resonance between Juliet and Portia.
}
\label{fig:hist2body_b}
\end{center}
\end{figure*}

\subsection{Two-body resonances between Uranian moons}
\label{sec:two_moon}

In Table \ref{tab:twobodystuff} we list computed properties of strong two-body 
first-order mean-motion resonances
in the inner Uranian satellite system.
We have computed characteristic libration frequencies for both resonance
terms (that corresponding to $\phi_{qi}$ and that corresponding 
to $\phi_{qj}$) for the $i$-th and $j$-th body in a $q-1:q$ resonance
and listed the maximum libration frequency 
\begin{equation}
\nu_{max} \equiv \max (\nu_i, \nu_j, \nu_{ei}, \nu_{ej}). \label{eqn:numax}
\end{equation}
Here libration frequencies $\nu_i, \nu_j$ are computed using equation \ref{eqn:omij}
and $\nu_{ei} , \nu_{ej}$ using equation \ref{eqn:nueij}.
We use semi-major axes and eccentricities from the beginning of the integration to perform these computations.
We identify which resonant term (that associated with $\phi_{qi}$ or $\phi_{qj}$) is larger
from the maximum libration frequency and this is also listed in Table \ref{tab:twobodystuff}.

Libration frequencies for the strongest first-order mean-motion resonances 
 are of order $10^{-7}$ Hz corresponding
to periods of $10^8$s (a few years).
By computing 
equation \ref{eqn:nueij_approx} from values for eccentricity and mass ratio 
listed in Table \ref{tab:stuff}
and inter-satellite separations in Table \ref{tab:twobodystuff}, and restoring units by 
multiplying by the  mean motion of the inner satellite (also listed in Table \ref{tab:stuff}),
we have checked that the approximation for the libration frequency (using equation \ref{eqn:nueij_approx})
 is within a factor of a few 
of the quantity more accurately calculated using  Laplace coefficients.

In Table \ref{tab:twobodystuff} we also list the ratio of eccentricity to critical eccentricity for the
stronger resonant subterm
\begin{equation}
e_m \equiv  \left\{ \begin{array}{c}
\frac{e_i}{e_{i,crit}} \\ 
\frac{e_j}{e_{j,crit}} 
\end{array}
{\rm for} 
\begin{array}{c} 
\max (\nu_i,\nu_{ei}) \ge \max (\nu_j, \nu_{ej}) \\
\max (\nu_i,\nu_{ei}) < \max (\nu_j, \nu_{ej}) 
\end{array}
\right.  \label{eqn:e_m}
\end{equation}
and these are computed using equation \ref{eqn:ecrit}.
Distance to resonance is estimated with the frequency
\begin{equation}
b_m \equiv {\min (|b_i|, |b_j|) }. \label{eqn:b_m}
\end{equation}
When $b_m/\nu_{max} \lesssim 1$ the pair of bodies is strongly influenced by the resonance.
It will be helpful later on to consider $b_m$ as a small divisor when we discuss three-body resonances in section \ref{sec:3per}.

The coefficients $b_i,b_j$ were computed
using equation \ref{eqn:AB_2} and with
precession rates calculated using equation 
\ref{eqn:ABoblate} (and so lacking contribution from secular satellite 
interactions). 
We use equation \ref{eqn:epsilonij} for $\epsilon_i, \epsilon_j$ to compute quantities such
as $\nu_{max}$ and $e_m$.

To compare the strengths of the $\phi_{qi}$ and $\phi_{qj}$ resonant terms we compute a ratio  
$\mu_m$
\begin{eqnarray} 
\mu_m \equiv \left\{ \begin{array}{l}\mu_{ij}\\ \mu_{ij}^{-1} \end{array}
  \qquad {\rm for} \qquad \begin{array}{c} \mu_{ij} <1 \\ \mu_{ij}>1\end{array} \right.\label{eqn:mum}
  \end{eqnarray}
with 
\begin{equation}
\mu_{ij} 
\equiv {\Gamma_j^{1 \over 2} \epsilon_j \over \Gamma_i^{1\over 2} \epsilon_j} = \left({\nu_{ej} \over \nu_{ei}} \right)^2.   
\label{eqn:muratio}
\end{equation}
corresponding to coefficients in the Hamiltonian, equation \ref{eqn:K4d}.
An energy for the dominant sub-term
\begin{equation}
\varepsilon_m \equiv \left\{  \begin{array}{l}  \epsilon_i \Gamma_i^{1\over 2}  \\
\epsilon_j \Gamma_j^{1 \over 2}   \end{array} \right. \qquad {\rm for} \qquad
 \begin{array}{c} \mu_{ij} <1 \\ \mu_{ij}>1\end{array} \label{eqn:varepsilon_m}
\end{equation}
is listed in Table \ref{tab:twobodystuff} divided by the energy for the dominant term
in the Cressida/Desdemona 46:47 resonance, denoted as $\varepsilon_{mCD}$.
We also compute the frequency ratio
\begin{equation}
\lambda_{olp} \equiv \left| {\dot \varpi_{ij} \over \nu_{max}} \right|, \label{eqn:lambdaoverlap}
\end{equation}
which is a parameter describing the proximity of the two resonance terms 
\citep{holman96,murray97}.

As can be seen from Table \ref{tab:twobodystuff},  and with the exception of
resonances involving Cupid, at the beginning of the integration
the bodies tend to be near but above the critical eccentricities for each resonance term.
Thus usually $\nu_{ei} > \nu_i$ and $\nu_{ej} > \nu_j$.
Cupid has a comparatively high eccentricity so $e_m >1$ for the 57:58 resonance with Belinda
and the 24:25 resonance with Perdita.

Only for the Cupid/Belinda 57:58, Belinda/Perdita 43:44 and Cressida/Desdemona 46:47 resonances is the system
clearly in the vicinity of resonance at the beginning of the integration with $b_m \lesssim 1$.
In the rightmost column in Table \ref{tab:twobodystuff} we compute 
this energy divided by that for the Cressida/Desdemona 46:47 resonance, allowing
a comparison of the relative energies of the resonant terms.  The energy in the Juliet/Portia
resonances is high because of the comparatively large masses of Juliet and Portia. 

\subsection{Intermittency in resonant angle histograms}
\label{sec:inter_hist}

Near a resonance, the resonant angle moves slowly or freezes. The distribution of
angle values measured in a  time interval peaks at the frozen angle and is not flat.
Examination of histograms of a resonant angle during different time intervals 
is a way to search for resonant interaction in a numerical integration.
For example,
a pair of bodies with a resonant angle librating about $\pi$  has an angle histogram that is strongly peaked at $\pi$.
If the  pair of bodies are distant from the resonance, then the angle circulates and the histogram would be flat.
Near a resonance separatrix, the histogram can peak at $\pi$ or 0 even if the angle circulates.

For each 500 data outputs (each spanning a time interval $5 \times 10^9$s long) 
in the numerical integration, we used orbital elements,
computed from the  state vectors, to create histograms
of the angles $\phi_{qi}$ and $\phi_{qj}$.   
These angles, modulo $2 \pi$, are binned in 18 angular bins.  
The result is a two-dimensional histogram, with time intervals along one axis and angle along the other.  
Each bin counts the number of times the angle  was in that angle bin during the time interval.
We note that sometimes
the sampling or data output period introduces structure into the histograms when the distribution should
be flat.  This happens when the angle plotted happens to have a period that is approximately an integer ratio
of the sampling period. When there are variations in the period of the angle, then such aliasing is rarer.
Unfortunately the integration output rate was not chosen with the creation of angle histograms in mind
so we cannot decrease the output period or resample it.

The structure-exhibiting resonant angle histograms for first-order mean-motion resonance angles involving two bodies
are shown in Figure \ref{fig:hist2body}
and Figure \ref{fig:hist2body_b}, 
with Figure \ref{fig:hist2body_b} focusing on first- and second-order resonances between Juliet and Portia. 
In Figure \ref{fig:hist2body}
when the color is black, the system spent no time with the resonant angle in that particular bin.
If the color is uniformly blue, then the angle was evenly distributed and was probably circulating.
When the angle remains fixed or librates about a particular value there is a peak in the histogram at this $y$ axis value.
The closest resonances, 
Cupid/Belinda 57:58, Belinda/Perdita 43:44 and Cressida/Desdemona 46:47 (at the top of Table \ref{tab:twobodystuff}
and with proximity measured as having a low value of $b_m/\nu_{max}$) 
have resonant angle histograms with particularly strong structure.
These pairs spend more time with the resonant angle near 0 or $\pi$.

Even though  Desdemona and Portia are not very near the 12:13 resonance (as seen from $b_m/\nu_{max}$ 
in Table \ref{tab:twobodystuff}), 
the resonant angle  $12 \lambda_{Des} - 13 \lambda_{Por} +\varpi_{Des}$ tends to remain near 0
and $12 \lambda_{Des} - 13 \lambda_{Por} + \varpi_{Por}$ spends more time near $\pi$.
Similarly $15\lambda_{Bia} - 16 \lambda_{Cres} + \varpi_{Cres} $ spends more time near $\pi$
than 0.  Figure \ref{fig:hist2body_b} shows that the 24:25 and 25:26 first-order resonances  between
Juliet and Portia could be important even though Juliet and Portia are nearer the weaker 49:51 
second-order mean-motion resonance.

Intermittent behavior is seen in the resonant angle histograms of the 57:58 resonance of Cupid and Belinda,
the 46:47 resonance of Cressida and Desdemona and the 43:44 resonance of Belinda and Perdita.  The angle
$57\lambda_{Cup} - 58\lambda_{Bel} - \varpi_{Cup}$ librates about 0 or $\pi$, making transitions between the two states.
Transitions between libration states are coupled
in the Cupid, Belinda and Perdita trio.  
For example, when the angle $43\lambda_{Bel} - 44\lambda_{Per} - \varpi_{Per}$ 
makes a transition from $\pi$ to 0 at $t\sim 8 \times 10^{11}$s
the angle 
$57\lambda_{Cup} - 58\lambda_{Bel} - \varpi_{Bel}$ makes a transition from 0 to $\pi$.
In contrast, Cressida and Desdemona's resonant angles undergo a variety of transitions 
but none of the other two-body angles in Figure \ref{fig:hist2body}  
 make transitions at the same time.

We could view the transitions of the resonant angles  as an example of `Hamiltonian intermittency' (e.g., \citealt{shev10}).
As discussed by \citet{shev10}, Hamiltonian intermittency is attributed to oscillations in the location of a separatrix or
sticky orbits (cantori) in the boundary of a chaotic layer.   Perhaps both mechanisms are possible here.
To investigate the source of chaotic behavior and associated intermittency we consider two
possible sources of chaotic behavior.  First, we consider the role of the
two resonant terms in an individual first-order mean-motion resonance, following \citet{holman96} who estimated
Lyapunov timescales in mean-motion resonances in the asteroid belt based on overlap between resonant subterms.
The Lyapunov exponents characterize the mean rate of exponential divergence 
of trajectories close to each other in the phase space. By Lyapunov timescale we mean
the inverse of the maximum Lyapunov exponent.
Second, in section \ref{sec:calc} we will discuss the Lyapunov timescale in resonant chains, 
when there are pairs of first-order mean motions resonances in trios of bodies. 

\subsection{Resonance overlap between subterms in individual first-order resonances}
\label{sec:resover}

If we can  compare our Hamiltonian model to 
the well-studied non-linear driven pendulum then we can estimate the Lyapunov timescale in it.
Because eccentricities are usually above or near the critical values we can assume that the system oscillates
about a mean eccentricity value.  In this case the coefficients of each resonant term are not strongly
dependent upon the variations in the momenta $\Gamma_i, \Gamma_j$.  
Using the strength ratio $\mu_m$, 
equation \ref{eqn:K4d} can be approximately transformed (via canonical transformation) to
\begin{eqnarray}
K (J,\phi; \Gamma, \varpi_{ij}) \approx 
{A\over 2} J^2 + b_r J  +  \Omega \Gamma  \qquad \qquad \qquad \qquad \\ 
 \qquad \qquad + \epsilon_m \left[ \cos \phi 
+ \mu_m \cos (\phi + \varpi_{ij})  
\right]\nonumber,
\end{eqnarray}
where $\phi$ is the angle $\phi_i$ or $\phi_j$ for the strongest term and is conjugate to
$J$.  The angle $\varpi_{ij}$ is conjugate to $\Gamma$ and $\Gamma$ is either $\Gamma_i$
or $\Gamma_j$ depending upon which resonant sub-term is dominant;
likewise the coefficient $b_r$ is either $b_i$ or $b_j$.
Here $\Omega$ is a perturbation frequency also representing the distance between the the two resonances;
$\Omega \sim \pm \dot \varpi_{ij}$.  The frequency of small oscillations for the dominant resonance
 $\nu_{max} = \sqrt{A \epsilon_m}$.

The Hamiltonian can be
recognized as a periodically-perturbed pendulum \citep{chirikov79,shev02,shev14} and
our description is equivalent to the forced-pendulum model for chaos in mean-motion
resonances in the asteroid belt by \citet{holman96,murray97}.
The periodically-perturbed pendulum exhibits chaotic behavior in the separatrix of the primary resonance.
Following \citet{chirikov79,shev02},
a unitless overlap parameter, $\lambda_{olp}$, can be constructed from the perturbation frequency
and frequency of small oscillations of the dominant resonance 
\begin{equation}
 \lambda_{olp}  = \frac{\Omega}{\nu_{max}} = \frac{\dot \varpi_{ij}}{\nu_{max}}.
\end{equation}
This parameter affects the separatrix width and the Lyapunov timescale inside the separatrix 
\citep{chirikov79,shev02,shev04,shev14}.
Whereas in the asteroid belt the separation between the two resonant subterms arises from secular
interactions with giant planets, here the separation arises from the oblateness of the planet.

We can use an approximation for the precession rate (equation \ref{eqn:dotvarpiapprox}) and
resonance libration frequencies (equation \ref{eqn:nu_ij_near}) for a closely-spaced system to
estimate 
\begin{equation}
\lambda_{olp} \sim 5.25 \max(m_i,m_j)^{-\frac{2}{3}} \delta_{ij}^\frac{7}{3} j_2 \left( \frac{R_p}{a_i} \right)^2  \label{eqn:lam_approx}
\end{equation}
where we have set $q \sim \delta_{ij}^{-1}$ for the nearest first-order mean-motion resonance.
The strong dependence on separation accounts for the differences in $\lambda_{olp}$ seen in 
Table \ref{tab:twobodystuff}.

Table \ref{tab:twobodystuff} shows that the 
perturbation strengths of the sub-term, $\mu_m$, are not small, so the
energy changes  due to the perturbation term each orbit 
 in the separatrix of the dominant resonance would be of order the energy in the resonance itself.   
However, inspection of Table \ref{tab:twobodystuff} shows that the overlap ratio $\lambda_{olp} \lesssim 0.1$
for most of the resonances.  This puts them in the regime described as  {\it adiabatic chaos} by 
\citet{shev08}.    
In this regime, the Lyapunov timescale for chaotic evolution is approximately
 the perturbation period  $T = 2 \pi/\Omega$
(logarithmically increasing only at very small $\lambda$, see equation 17 by 
\citet{shev08}).  In units of the resonance libration period 
the Lyapunov timescale is approximately inversely proportion to $\lambda$.
As the resonance libration periods are of order 1-10 years 
(frequencies are listed in Table \ref{tab:twobodystuff}), and the 
overlap parameters $\lambda_{olp} \lesssim 0.1$,
the Lyapunov timescale would be in the regime of 10-100 years.
The overlap of these resonant subterms
might account for some of the  intermittency present in the resonant angles during the integration.  
We note that
the separatrix width, in units of energy, depends on $\lambda_{olp}^2$ and is small when $\lambda_{olp}<1$
(the $W$ parameter $\propto \lambda_{olp}^2$;
equation 5 of \citet{shev08}, and the separatrix width is equal to this energy, see figure 1 of \citet{shev04}).
Consequently the volume of phase space in which chaotic diffusion takes place is small in the adiabatic regime.
Only for the more widely-spaced bodies is the overlap parameter in a regime giving a comparatively short
Lyapunov timescale and a significant width in the chaotic region associated with the resonance separatrix.

Can we learn anything from considering what happens near a spherical planet or with $J_2=0$?
Equation \ref{eqn:lam_approx} implies that $\lambda_{olp} \to 0$ in this limit and we would expect
integrable mean-motion resonances (and so no chaotic behavior).  In contrast,
\citet{duncan97} found that an integration with $J_2=0$ exhibited more instability and had 
a shorter crossing timescale, opposite to what we expect.
We have neglected the role of secular interaction terms between bodies, 
and when $J_2 \to 0$ perhaps secular interactions
between distant moons become more important.

The overlap of sub-terms in individual mean-motion resonances, particularly important for pairs 
of bodies that are not the nearest ones,  
  could account for
transitions of a single resonant angle from a state near 0 to $\pi$ and vice versa.  However,
this mechanism  would not
account for coupled variations in angles in pairs of bodies, or coupled variations in semi-major axis between
more than two bodies.  Since numerical integrations have shown that integration of fewer
moons  can increase the crossing timescale \citep{french12},
we are also interested in mechanisms involving additional moons for the intermittency
 in the resonant angles.

\begin{table*}
\begin{minipage}{166mm}
\vbox to 190mm{\vfil
\caption{\large Properties of strong first-order two-body resonances \label{tab:twobodystuff} }
\begin{tabular}{@{}ll ll crrrrrrrr}
\hline
(1)           &   (2)            & (3)     & (4)        &(5)                   &(6) & (7) &(8) & (9) & (10) &(11)  &(12) &(13)\\ 
$i$           &   $j$           & q-1:q &   $\delta_{ij}$ &$\phi$  &$\nu_{max}$(Hz) & $\frac{\nu_{max}}{n_i}$ & $\frac{b_m}{\nu_{max}}$  & $\frac{b_m}{n_i}$ & $e_m$ & $\lambda_{olp}$ & $\mu_\epsilon$ & $\frac{\varepsilon_m}{\varepsilon_{mCD}}$ \\
\hline
 Cressida & Desdemona & 46:47 &   0.014 & $\phi_j$ & 1.1e-07 & 6.9e-04 &    1.4 & 9.6e-04 &   0.89 &  0.065 &   1.28 &    1.00  \\ 
  Belinda &   Perdita & 43:44 &   0.015 & $\phi_j$ & 2.1e-07 & 1.8e-03 &    0.1 & 1.3e-04 &  10.50 &  0.018 &   6.49 &    0.80  \\ 
    Cupid &   Belinda & 57:58 &   0.012 & $\phi_i$ & 2.2e-07 & 1.8e-03 &   -2.4 & -4.5e-03 &   5.51 &  0.013 &   0.09 &   -0.17  \\ 
Desdemona &    Portia & 12:13 &   0.055 & $\phi_i$ & 4.5e-08 & 3.0e-04 &    9.1 & 2.7e-03 &   0.29 &  0.508 &   0.36 &   -1.29  \\ 
   Bianca &  Cressida & 15:16 &   0.044 & $\phi_i$ & 3.1e-08 & 1.8e-04 &    8.6 & 1.6e-03 &   2.37 &  0.748 &   0.55 &   -0.65  \\ 
 Rosalind &   Perdita &  7: 8 &   0.093 & $\phi_j$ & 1.2e-08 & 9.1e-05 &  -20.7 & -1.9e-03 &   7.38 &  2.083 &   4.78 &    0.08  \\ 
Desdemona &    Juliet & 24:25 &   0.027 & $\phi_j$ & 7.6e-08 & 4.9e-04 &  -29.8 & -1.5e-02 &   3.80 &  0.160 &   0.54 &    3.73  \\ 
 Cressida &    Juliet & 16:17 &   0.042 & $\phi_j$ & 4.4e-08 & 2.8e-04 &   62.7 & 1.8e-02 &   2.56 &  0.429 &   0.69 &    3.96  \\ 
    Cupid &   Perdita & 24:25 &   0.027 & $\phi_j$ & 1.6e-08 & 1.3e-04 &  -93.9 & -1.2e-02 &  67.58 &  0.424 &   0.58 &    0.00  \\ 
   Bianca & Desdemona & 11:12 &   0.059 & $\phi_i$ & 1.7e-08 & 9.9e-05 &  -92.9 & -9.2e-03 &   2.60 &  1.818 &   0.70 &   -0.30  \\ 
   Portia &  Rosalind & 11:12 &   0.058 & $\phi_j$ & 4.0e-08 & 2.8e-04 &  -95.3 & -2.7e-02 &   0.41 &  0.505 &   2.78 &    1.87  \\ 
   Portia &   Perdita &  4: 5 &   0.156 & $\phi_j$ & 1.6e-08 & 1.1e-04 & -192.5 & -2.1e-02 &   3.18 &  2.848 &  13.09 &    0.36  \\ 
 Rosalind &     Cupid & 10:11 &   0.064 & $\phi_j$ & 1.4e-08 & 1.1e-04 & -223.4 & -2.3e-02 &   3.92 &  1.310 &   8.22 &    0.02  \\ 
   Portia &     Cupid &  5: 6 &   0.126 & $\phi_j$ & 1.5e-08 & 1.1e-04 & -227.2 & -2.4e-02 &   1.63 &  2.541 &  22.66 &    0.07  \\ 
   Juliet &     Cupid &  4: 5 &   0.156 & $\phi_j$ & 7.5e-09 & 5.1e-05 & -440.5 & -2.2e-02 &   2.03 &  6.548 &  14.97 &    0.03  \\ 
 Rosalind &   Belinda &  9:10 &   0.076 & $\phi_i$ & 1.2e-08 & 9.0e-05 &  492.8 & 4.4e-02 &   0.67 &  1.785 &   0.74 &   -0.34  \\ 
   Portia &   Belinda &  5: 6 &   0.139 & $\phi_j$ & 1.4e-08 & 9.8e-05 &  634.4 & 6.2e-02 &   0.28 &  2.947 &   2.04 &    1.29  \\ 
\hline
  Belinda &   Perdita & 42:43 &   0.015 & $\phi_j$ & 2.0e-07 & 1.7e-03 &  -13.0 & -2.3e-02 &  10.33 &  0.018 &   6.49 &    0.80  \\ 
  Belinda &   Perdita & 44:45 &   0.015 & $\phi_j$ & 2.1e-07 & 1.8e-03 &   12.7 & 2.3e-02 &  10.68 &  0.018 &   6.49 &    0.79  \\ 
 Cressida & Desdemona & 47:48 &   0.014 & $\phi_j$ & 1.1e-07 & 6.9e-04 &   32.1 & 2.2e-02 &   0.91 &  0.064 &   1.28 &    1.00  \\ 
 Cressida & Desdemona & 45:46 &   0.014 & $\phi_j$ & 1.1e-07 & 6.8e-04 &  -30.0 & -2.0e-02 &   0.88 &  0.066 &   1.28 &    1.00  \\ 
Desdemona &    Juliet & 25:26 &   0.027 & $\phi_j$ & 7.8e-08 & 5.1e-04 &   48.4 & 2.5e-02 &   3.92 &  0.154 &   0.54 &    3.70  \\ 
Desdemona &    Portia & 11:12 &   0.055 & $\phi_i$ & 4.3e-08 & 2.8e-04 & -264.0 & -7.4e-02 &   0.28 &  0.534 &   0.36 &   -1.30  \\ 
   Bianca &  Cressida & 14:15 &   0.044 & $\phi_i$ & 2.9e-08 & 1.7e-04 & -351.0 & -6.1e-02 &   2.25 &  0.796 &   0.54 &   -0.66  \\ 
 Cressida &    Juliet & 15:16 &   0.042 & $\phi_j$ & 4.2e-08 & 2.7e-04 & -157.5 & -4.2e-02 &   2.45 &  0.455 &   0.69 &    4.00  \\ 
    Cupid &   Belinda & 56:57 &   0.012 & $\phi_i$ & 2.2e-07 & 1.8e-03 &  -11.9 & -2.2e-02 &   5.44 &  0.014 &   0.09 &   -0.17  \\ 
    Cupid &   Belinda & 58:59 &   0.012 & $\phi_i$ & 2.2e-07 & 1.9e-03 &    6.8 & 1.3e-02 &   5.58 &  0.013 &   0.09 &   -0.17  \\ 
   Portia &  Rosalind & 12:13 &   0.058 & $\phi_j$ & 4.2e-08 & 2.9e-04 &  185.1 & 5.4e-02 &   0.44 &  0.483 &   2.79 &    1.85  \\ 
   Juliet &    Portia & 24:25 &   0.027 & $\phi_i$ & 1.2e-07 & 8.2e-04 &  -22.6 & -1.9e-02 &   1.30 &  0.091 &   0.67 &  -28.60  \\ 
   Juliet &    Portia & 25:26 &   0.027 & $\phi_i$ & 1.2e-07 & 8.4e-04 &   24.8 & 2.1e-02 &   1.34 &  0.089 &   0.67 &  -28.38  \\ 
   Juliet &    Portia & 26:27 &   0.027 & $\phi_i$ & 1.3e-07 & 8.5e-04 &   70.3 & 6.0e-02 &   1.38 &  0.087 &   0.67 &  -28.14  \\ 
 Rosalind &   Belinda &  8: 9 &   0.076 & $\phi_i$ & 1.1e-08 & 8.4e-05 & -715.2 & -6.0e-02 &   0.62 &  1.907 &   0.74 &   -0.34  \\ 
Desdemona &  Rosalind &  6: 7 &   0.116 & $\phi_j$ & 5.6e-09 & 3.6e-05 & 1758.1 & 6.4e-02 &   0.85 &  7.760 &   1.00 &    0.13  \\ 
\hline
\end{tabular} 
{\\ The properties of strong first-order mean-motion resonances in
the Uranian satellite system.
Columns:  1,2 satellite names corresponding to bodies $i,j$.
The resonant arguments are $\phi_{qi} = q \lambda_j + (1-q) \lambda_i - \varpi_i$ and 
$\phi_{qj} =q \lambda_j + (1-q) \lambda_i - \varpi_j$.
Col 3: The integers $q-1:q$. 
Col 4: The spacing between the two bodies $\delta_{ij}$ computed using equation \ref{eqn:deltaij}. 
Col 5:  The dominant resonant argument (that with larger libration frequency)
 is denoted as $\phi_i$ if the $\phi_{qi}$ angle is important
or $\phi_j$ if the $\phi_{qj}$ angle is important.
Col 6: The frequency of librations in resonance, $\nu_{max}$, (equation \ref{eqn:numax})  in units of Hz.
This frequency is computed using equations \ref{eqn:omij},\ref{eqn:nueij}) and \ref{eqn:epsilonij}.
Col 7: $\nu_{max}$ divided by the innermost body's mean motion, $n_i$.
Col 8:  The distance to resonance $b_m$ (equation \ref{eqn:b_m}) in units of $\nu_{max}$.
When $|b_m/\nu_{max}| \lesssim 1$ the system is near resonance.  
Here the frequencies $b_i,b_j$ are computed using equation \ref{eqn:AB_2} and $b_m$,
the distance to the dominant resonant argument.
Col 9:  The distance to resonance in units of the innermost body's mean motion or the ratio $b_m/n_i$.
Col 10:
The ratio of initial eccentricity to critical eccentricity for the dominant argument (see
equation \ref{eqn:e_m},   and this is computed using equation \ref{eqn:ecrit}).  
Col 11:
The unitless overlap ratio, $\lambda_{olp}$,  (equation \ref{eqn:lambdaoverlap}) 
describing the proximity of the $\phi_{qi}$ and $\phi_{qj}$ resonances.   
Col 12: The unitless parameter $\mu_\epsilon$, 
 the ratio of $\phi_{qi}$ vs $\phi_{qj}$ resonance strengths (see equation \ref{eqn:mu_e}).
Col 13: Energy of the argument $\varepsilon_m$ (equation \ref{eqn:varepsilon_m}) divided by that for
the Cressida/Desdemona 46:47 resonance.
The resonances have been divided into two groups.  For each satellite pair, the top set lists only the nearest first-order resonance.
The bottom set includes more-distant resonances.
\\
}}
\end{minipage}
\end{table*}

\section{Three-body interactions}
\label{sec:search3}

Overlap of three-body multiplets is a source of chaos in the asteroid belt \citep{nesvorny98,murray98}.
\citet{quillen11} proposed that three-body resonances were responsible for slow, chaotic diffusion in the semi-major
axes of bodies in integrated planar closely-packed multiple-planet systems.
Three-body resonances in the Uranian satellite system may account for some of 
the coupled variations we see  between three or more bodies.
To explore this possibility,
 we searched the inner Uranian satellite system for strong three-body resonances.
When a three-body resonance is strong,
 the associated Laplace angle freezes or librates \citep{nesvorny98,nesvorny98b,smirnov13}.
 We search for time periods when Laplace angles are slowly moving  
 and then discuss comparisons between histograms of
resonant angles and variations in orbital elements between trios of bodies.

\subsection{Searching for nearby three-body resonances}

The three-body resonances discussed by \citet{quillen11}
are specified by two integers $p,q$.
The  p:-(p+q):q resonance 
is associated with a Laplace angle 
\begin{equation}
\theta = p \lambda_i - (p+q)\lambda_j + q \lambda_k \label{eqn:theta}
\end{equation}
that involves mean longitudes of three bodies $i,j,k$ 
where we assume that the semi-major axes $a_i< a_j <a_k$.
The Laplace angle is slowly moving when the frequency
\begin{equation}
\dot \theta \approx p n_i - (p+q) n_j + q n_k \sim 0 \label{eqn:dottheta}
\end{equation}
with $n_i, n_j, n_k$ the mean motions of the three bodies.

\begin{figure*}
\begin{center}
\includegraphics[width=15cm]{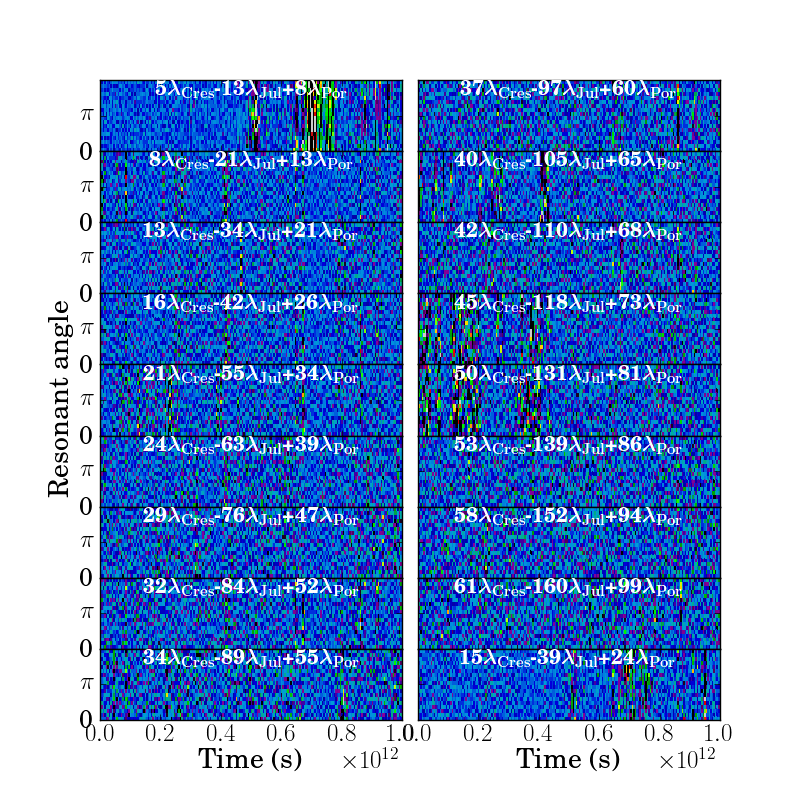}
\caption{Histograms of resonant angles of nearby three-body resonances
for Cressida, Juliet and Portia.
The resonant angle plotted is labelled in each panel.
}
\label{fig:CresJulPor}
\end{center}
\end{figure*}

\begin{figure*}
\begin{center}
\includegraphics[width=15cm]{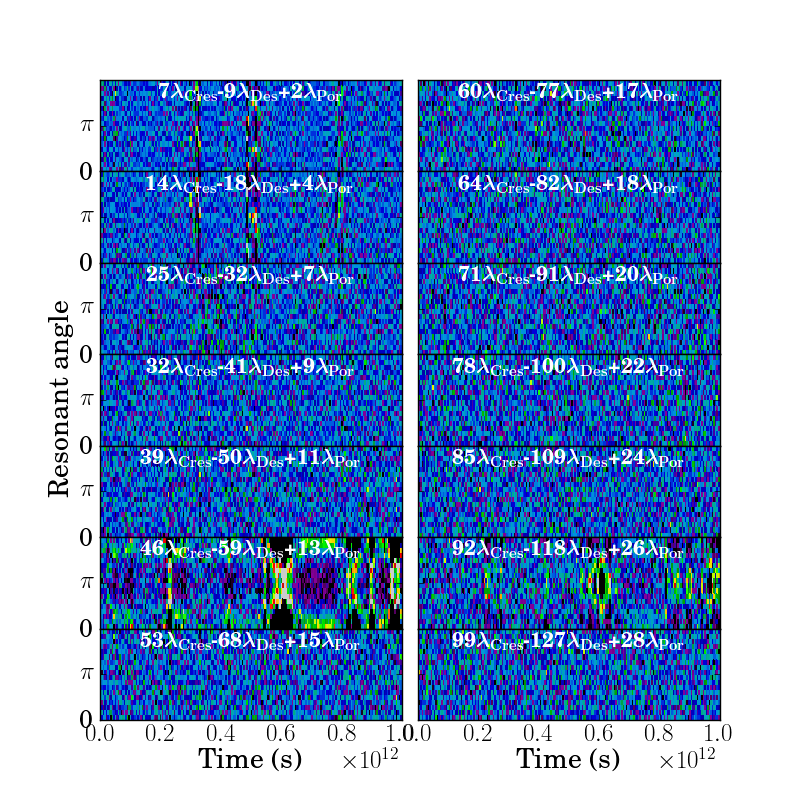}
\caption{Histograms of resonant angles of nearby three-body resonances
Cressida, Desdemona and Portia.
}
\label{fig:CresDesPor}
\end{center}
\end{figure*}

For trios of bodies,
we searched for integers $p,q$  that minimized $|\dot \theta|$.
For the trios Cressida, Juliet and Portia and Cressida, Desdemona and Portia
we list  three-body resonant angles, with $|\dot \theta| < 6 \times 10^7$~Hz at some
time in the interval $t=0$--$10^{12}$s,
in Table \ref{tab:3dist}, and we plot histograms of these resonant angles
in Figures \ref{fig:CresJulPor} and \ref{fig:CresDesPor}.
We limited our search to $p,q<100$ as Laplace coefficients (and so resonant strengths) 
are truncated exponentially  
with $p \delta_{ij}>1$ or $q\delta_{jk}>1$, with $\delta_{ij},\delta_{jk}$ describing 
the distances between the moons (\citealt{quillen11}, 
and as shown in equation \ref{eqn:approx}).

Gravitational interactions only involve two bodies, and it is only via canonical transformation
that we derive a Hamiltonian that contains a three-body Laplace angle.
\citet{quillen11} estimated three-body resonance strengths assuming that the dominant contribution was
from two zeroth-order (in eccentricity) perturbation terms,
\begin{equation}
W_{ij,p} \cos p(\lambda_i - \lambda_j)  + W_{ij,q} \cos q(\lambda_j - \lambda_k),
\end{equation}
that are Fourier components of two-body interaction terms. 
A near-identity canonical transformation gives a 
Hamiltonian in the vicinity of three-body resonance lacking these two terms
\begin{multline}
H(\vec \Lambda, \vec \lambda) = \sum_{l=i,j,k} - \frac{m_l^3}{2 \Lambda_i^2} \\
+ \epsilon_{pq} 
\cos ( p \lambda_i  - (p+q) \lambda_j +  q \lambda_k). \label{eqn:H3}
\end{multline}
The coefficient $\epsilon_{pq}(\vec \Lambda)$ is  sensitive to divisors $n_{ij}$ and $n_{jk}$
that are the difference in mean motions of the two bodies (see equation 23 for $\epsilon_{pq}$ of \citealt{quillen11}) and
can be considered a second-order perturbation (and depending on a higher power of moon mass) as it involves
a product of the coefficients $W_{ij,p}$ and $W_{jk,q}$.
The dependence on divisors  $n_{ij}$ and $n_{jk}$
suggests that all the resonances listed in Table \ref{tab:3dist} should have similar strengths.
However, we can see by comparing the resonant angle histograms in Figures \ref{fig:CresJulPor} and \ref{fig:CresDesPor}
that this is probably not the case.

We first check to see if the resonant angles freeze only if the three bodies are very
near resonance. 
For the Cressida, Juliet and Portia trio there is a time when the bodies are very near the 29:-76:47 
resonance (with $|\dot \theta| < 10^{-10}$ Hz, as listed in Table \ref{tab:3dist}).  Most of the other resonances have 
 minimum distance $|\dot \theta |\sim 10^{-7}$ Hz.  
 Despite proximity to resonance, the 29:-76:47  resonant angle does not show
 more structure than the other angles in Figure \ref{fig:CresJulPor}. 
 The Cressida, Desdemona and Portia trio  is near both the 39:-50:11 and 46:-59:13 resonances
 but only the 46:-59:13  resonant angle shows strong structure in Figure \ref{fig:CresDesPor}.
 We find that
 proximity is not the only factor governing three-body resonant strength (as inferred through structure
 in a resonant angle histogram).
 
As discussed in section \ref{sec:twobody},  Cressida and Desdemona
are near or in the 46:47 first-order mean-motion resonance and Desdemona and Portia 
are near their 12:13 
first-order mean-motion resonance.
The two resonant angles from the nearby first-order mean-motion resonances are
\begin{eqnarray}
\phi_{p} &=& 47 \lambda_{Des} - 46 \lambda_{Cres} - \varpi_{Des} \nonumber \\
\phi_{q} &=& 13  \lambda_{Por} - 12 \lambda_{Des} - \varpi_{Des} \nonumber 
\end{eqnarray}
and the difference between these angles
\begin{eqnarray}
\theta &=& \phi_{q} - \phi_{p}  \nonumber \\
&=& 46  \lambda_{Cres} - 59 \lambda_{Des} +  13  \lambda_{Por}
\end{eqnarray}
and equivalent to the  46:-59:13 Laplace angle involving the three bodies 
Cressida,  Desdemona and Portia.
This particular three-body resonance could be strong because each consecutive pair of
bodies is near a first-order mean-motion resonance.
We describe this setting as a `resonant chain'.
The 39:-50:11 three-body resonance, 
perhaps because it is not near any first-order mean-motion resonances between pairs of bodies,
is weaker than the 46:-59:13 resonance.
In Figure \ref{fig:CresDesPor} the 92:-118:26 angle histogram also shows structure, 
however this angle is a multiple of two of the 46:-59:13 Laplace angle.
The 92:-118:26 Laplace angle histogram may show structure due to the 46:-59:13
three-body resonance.

In Figure \ref{fig:CresJulPor} the  5:-13:8 angle histogram shows structure suggesting that this
resonance with Cressida, Juliet and Portia might be stronger 
than the other three-body resonances in this trio. 
If Cressida, Juliet and Portia  are near the 5:-13:8 resonance then they are also near resonances described with 
integer multiples of this, the 10:-25:16 (multiply by 2)
and the 15:-39:24 (multiply by 3) resonances.   For resonance strengths estimated from the zeroth-order interaction 
terms alone, the resonance strength energy coefficient
$\epsilon_{pq} \sim \epsilon_{2p,2q}$ and so on for other multiples
as long as the strength is not 
exponentially truncated by the Laplace coefficients.
The 5:-13:8 three-body resonance may be strong because of the contribution from higher-index multiples.

Is the 5:-13:8 resonance with Cressida, Juliet and Portia  also near two two-body first-order resonances and a Laplace angle
associated with a resonant chain?
As seen in Table \ref{tab:twobodystuff}  Cressida and Juliet are fairly near the 15:16 first-order resonance
and Juliet and Portia fairly near the 23:24 first-order resonance.  
The 15:-39:24 Laplace angle is a multiple of 3 times the 5:-13:8 Laplace angle.
 The  5:-13:8  Laplace angle
may show structure due to the 15:16 resonance between  Cressida and Juliet or the 23:24 resonance
between Juliet and Portia.  The histogram on the lower right in Figure \ref{fig:CresJulPor} shows the
the histogram for the Laplace angle 15:-39:24 with $\dot \theta = 1.8\times 10^{-6}$~Hz, and this angle shows  structure
even though the distance to resonance is larger than the other considered
Laplace angles.  
The structure in the 5:-13:8 Cressida, Juliet and Portia angle histogram could be explained by the combined effects
of the 5:-13:8 and multiples of this resonance, each with strength contributed with
zeroth-order terms, or because the 15:-39:24 resonance is near a chain
of first-order resonances.

\begin{table}
\vbox to 120mm{\vfil
\caption{\large Potential three-body resonances \label{tab:3dist} }
\begin{tabular}{@{}lrlrlr}
\hline
\multicolumn{2}{c}{Cres/Jul/Por} & \multicolumn{2}{c}{Cres/Des/Por} \\
$p$:-($p$+$q$):$q$ & $\dot \theta$(Hz) & $p$:-($p$+$q$):$q$ & $\dot \theta$(Hz) \\
\hline
5:-13:8     &  6.0e-07  & 7:-9:2     & -2.4e-07 \\
8:-21:13   & -1.2e-07 & 14:-18:4 & -4.8e-07 \\ 
13:-34:21 &  3.9e-07 & 25:-32:7 & 5.3e-07 \\ 
16:-42:26 & -2.3e-07 & 32:-41:9 & 2.0e-07 \\ 
21:-55:34 &  1.9e-07 & 39:-50:11 & -1.8e-10 \\ 
24:-63:39 & -3.5e-07 & 46:-59:13 & -7.9e-11\\ 
29:-76:47 & -5.4e-11 & 53:-68:15 &-1.3e-07\\ 
32:-84:52 & -4.7e-07 & 60:-77:17 &-3.7e-07\\ 
34:-89:55 &  5.8e-07 &64:-82:18 &4.0e-07\\ 
37:-97:60 & -4.0e-11 &71:-91:20 &7.5e-08\\ 
40:-105:65 & -5.8e-07 & 78:-100:22 &-3.6e-10\\
42:-110:68 & 3.8e-07 & 85:-109:24 &8.9e-11\\ 
45:-118:73 & -2.4e-10 & 92:-118:26 &-1.6e-10\\ 
50:-131:81 & 1.8e-07 & 99:-127:28 &-2.7e-08 \\ 
53:-139:86 & -4.7e-08 & & \\ 
58:-152:94 & -1.1e-10 & &\\ 
61:-160:99 & -1.6e-07 & &\\
\hline
\end{tabular} 
{\\ The first and third columns list $p$:-($p$+$q$):$q$ with $p,q<100$, such that the
frequency $\dot \theta = p n_i - (p+q)n_j + q n_k$ has $|\dot \theta| < 6 \times 10^7$~Hz at some
time in the integration with $t<10^{12}$s.
The second and fourth columns list $\dot \theta$ in Hz.   The three bodies are 
Cressida, Juliet and Portia for the left two columns and Cressida, Desdemona and Portia for the right two columns. 
Histograms of the resonant angles are shown in Figures \ref{fig:CresJulPor} and \ref{fig:CresDesPor}.
}}
\end{table}

\subsection{Comparing variations in angle histograms with variations in orbital elements}

\begin{figure*}
\begin{center}
\includegraphics[width=15cm]{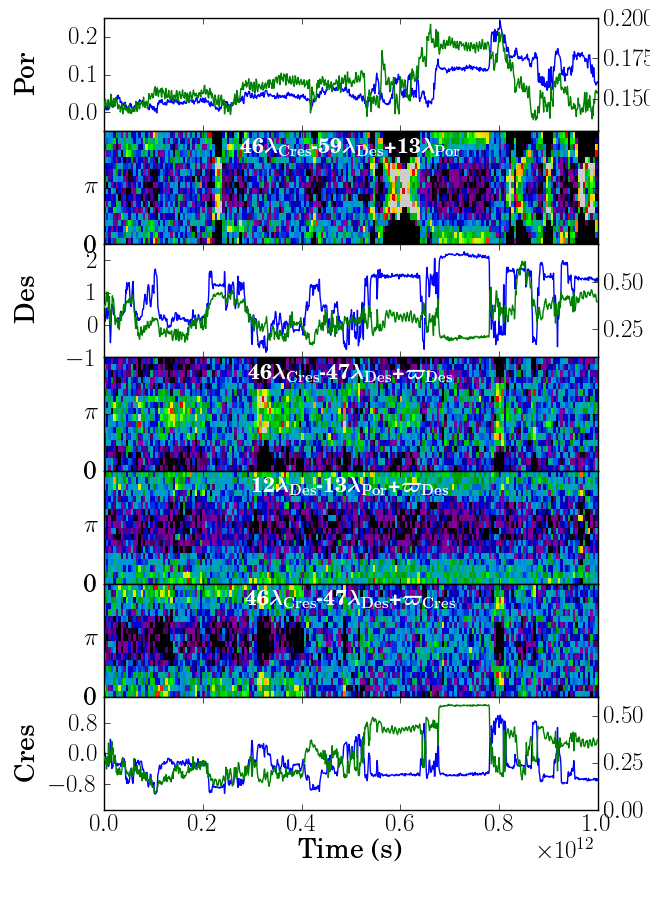}
\caption{Two- and three-body resonances influencing Cressida, Desdemona and Portia.
We plot both resonant angles (as histograms), semi-major axes (blue lines) and eccentricities (green lines)
so that they can be directly compared.   
Scaling for semi-major axes and eccentricities is the same as in Figure \ref{fig:geom_ae}.
Transitions in the 46:-59:13 Laplace angle with  Cressida, Desdemona and Portia
are coincident with coupled variations in semi-major axes of the three moons.
}
\label{fig:combo_a}
\end{center}
\end{figure*}

To explore the role of three-body angles
we  compare the structure seen in histograms  of  two-body and
three-body resonant angles with variations in orbital elements. 
The strongest structure seen in the histogram of a Laplace
angle was that seen in the  46:-59:13 angle with Cressida, Desdemona and Portia.
We plot in Figure \ref{fig:combo_a} the 46:-59:13  Laplace angle histogram, the resonant angle
histograms
for the 46:47 first-order resonance between Cressida and Desdemona,
the 12:13 resonance between Desdemona and Portia
and semi-major axes and eccentricities  for the three bodies as a function of time.
We find that transitions between states in the three-body resonant angle
are simultaneous with variations in semi-major axis in all three bodies.
The transitions in the three-body resonant angles are more important than
those seen in the two-body resonant angles.
For example, at $t \sim 3.5 \times 10^{11}$s the angle $46\lambda_{Cres} - 47\lambda_{Des} + \varpi_{Cres}$
flips from 0 to $\pi$ and there are only weak variations in $a_{Cres},a_{Des}$
at this time.  However at $t \sim 4 \times 10^{11}$s the Laplace angle
$46\lambda_{Cres} - 47\lambda_{Des} + 13 \lambda_{Cres}$ varies from 0 to $\pi$
and coupled variations in semi-major axis of all three bodies are seen.
Cressida and Portia move inward as Desdemona moves outward, as predicted
from conserved quantities present when a three-body resonance is important \citep{quillen11}.  
Transitions of the Laplace angle are better associated with jumps in semi-major axis
of all three bodies than the transitions in the two-body resonant angles.

Coupled motions in the semi-major axes of three bodies arise from a Hamiltonian that
contains a three-body Laplace angle. 
Using Hamilton's equation on equation \ref{eqn:H3}
\begin{eqnarray}
\dot \Lambda_i  = -\frac{\partial H}{\partial \lambda_i} = p \epsilon_{pq} 
\sin ( p \lambda_i  - (p+q) \lambda_j +  q \lambda_k). 
\end{eqnarray}
If the Laplace angle is quickly circulating then on average $\Lambda_i$ (the Poincar\'e coordinate dependent on $a_i$)
does not change.
However if  the Laplace angle remains fixed at $\pi/2$ then $\Lambda_i$ can increase or decrease, depending
on the sign of $\epsilon_{pq}$.
By similarly computing $\dot \Lambda_j$ and $\dot \Lambda_k$ we find that
simultaneous variation in the semi-major axis of the three bodies would take place with the inner and outer bodies moving
together and the middle one moving in the opposite direction.

\begin{figure*}
\begin{center}
\includegraphics[width=15cm]{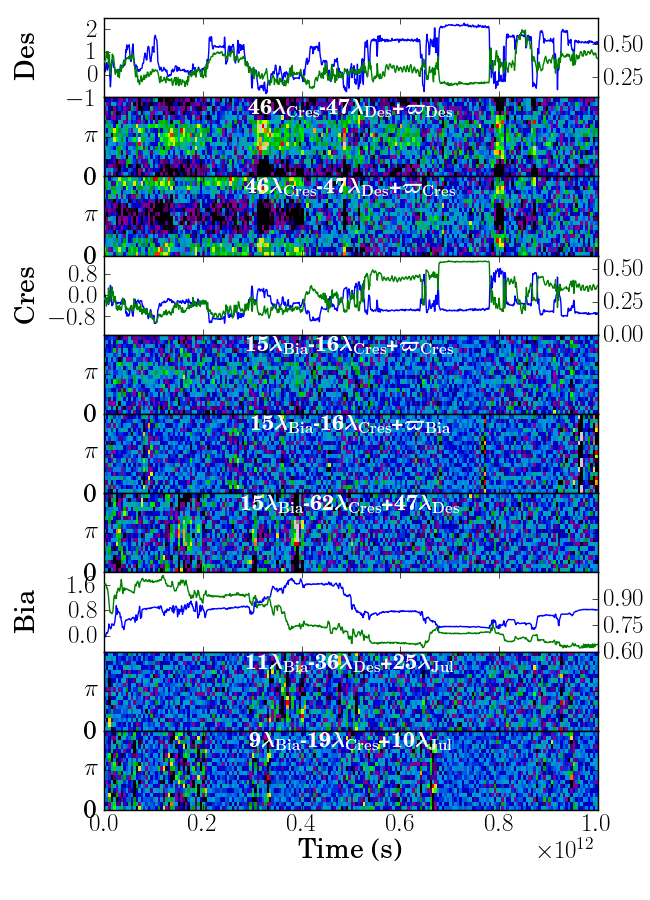}
\caption{Two- and three-body resonances influencing Bianca, Cressida and Desdemona. 
We plot both resonant angles (histograms), semi-major axes (blue lines) and eccentricities (green lines)
so that they can be directly compared.
Scaling for semi-major axes and eccentricities is the same as in Figure \ref{fig:geom_ae}.
Variations in the semi-major axis of Bianca tend to happen during transitions in three-body Laplace angles. 
}
\label{fig:combo_b}
\end{center}
\end{figure*}

In Figure \ref{fig:combo_b} we plot resonant angles and orbital elements with the goal of understanding
the variations in Bianca's orbit.
A three-body resonance influencing Bianca appears to be
the 15:-62:47 between Bianca, Cressida and Desdemona; it is
in proximity to the 15:16 first-order mean-motion resonance between Bianca and Cressida
and the 46:47 first-order mean-motion resonance between Cressida and Desdemona.
This is a resonant chain.
The 11:-36:25 resonance between Bianca, Desdemona and Juliet maybe responsible
for variations in Bianca's orbital elements at $t \sim 3.5-5 \times 10^{11}$s.
This is near the 11:12 first-order mean-motion resonance between Bianca and Desdemona and
the 24:25 first-order mean-motion resonance between Desdemona and Juliet, so it too is a resonant chain.
The 9:-19:10 resonance between Bianca, Cressida and Juliet is not near any two-body resonances,
and neither is it a multiple of the Laplace angle of a resonant chain.
Since it has low $p,q$ it may be strong because resonances associated with multiples of the 
resonant angle contribute to its strength.
Most of the variations in Bianca's semi-major axis are correlated with periods of
time where three-body Laplace angles are slowly moving or undergoing transitions.

\begin{figure*}
\begin{center}
\includegraphics[width=15cm]{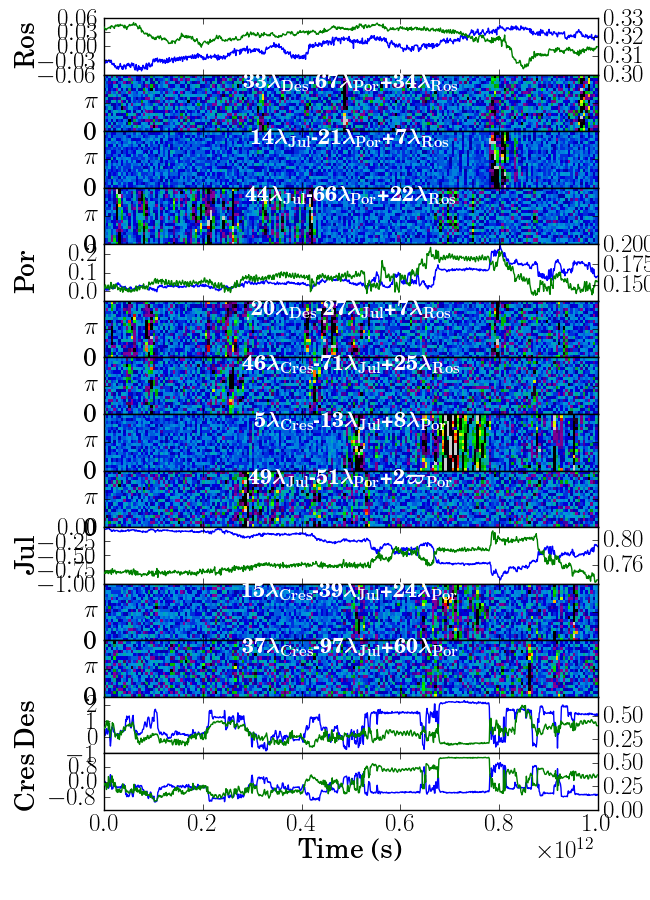}
\caption{Two- and three-body resonances influencing Desdemona, Juliet, Portia and Rosalind.
We plot both resonant angles (histograms), semi-major axes (blue lines) and eccentricities (green lines)
so that they can be directly compared.
Scaling for semi-major axes and eccentricities is the same as in Figure \ref{fig:geom_ae}.
}
\label{fig:combo_d}
\end{center}
\end{figure*}

In Figure \ref{fig:combo_d} we show additional angle histograms linking 
motions of Desdemona, Juliet, Portia and Rosalind.
Not all variations in orbital elements are explained.  For example, Rosalind drops
in eccentricity at $t \sim 8.5 \times 10^{11}$s without any strong change in semi-major axis. 
This could be due to a secular resonance that we have not identified.
A small jump in Rosalind's semi-major axis at $t \sim 4 \times 10^{11}$s is most likely
due to a Desdemona, Juliet and Rosalind coupling such as the 20:-27:7 resonance
as  Desdemona and Rosalind both move outwards
while Juliet moves inward.  
%
The 20:-27:7 resonance of Desdemona, Juliet and Rosalind is a resonant chain but not
with consecutive pairs; rather, the chain involves
the 6:7 first-order resonance between Desdemona and Rosalind (the outer two bodies) 
and the 20:21 between Desdemona and Juliet. 
Juliet, Portia and Rosalind are near a 2:-3:1 Laplace resonance that could be strong
because many of its multiples would contribute to the resonance. 

\begin{figure*}
\begin{center}
\includegraphics[width=15cm]{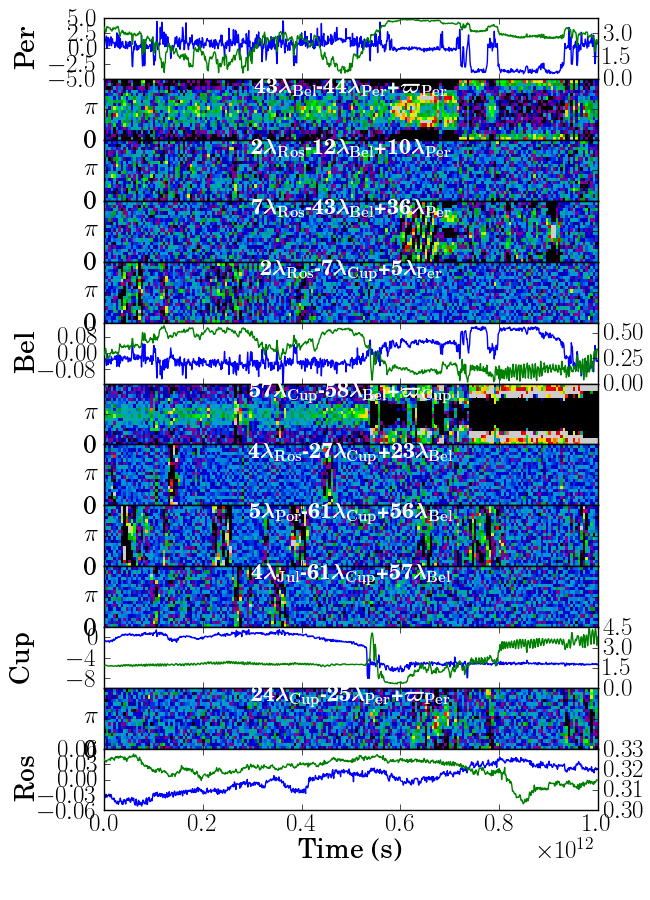}
\caption{Two- and three-body resonances influencing Rosalind, Cupid, Belinda and Perdita.
We plot both resonant angles (histograms), semi-major axes (blue lines) and eccentricities (green lines)
so that they can be directly compared.
Scaling for semi-major axes and eccentricities is the same as in Figure \ref{fig:geom_ae}.
The 57:58 Cupid/Belinda and 24:25 Cupid/Perdita two-body resonances account for many 
of the variations in orbital elements.  The presence of three-body resonances involving Portia or Juliet
with Cupid may account for the sensitivity of Cupid's crossing timescale to the presence of these bodies.
}
\label{fig:combo_c}
\end{center}
\end{figure*}

In Figure \ref{fig:combo_c} we examine variations in Cupid, Belinda and Perdita.
The  two-body first-order resonances, the  57:58 between Cupid and Belinda, the  24:25 between Cupid and Perdita 
and the 43:44 between Belinda and Perdita  account for many of the variations
in orbital elements.  
However, a number of three-body angles show structure.
The 7:-43:36 Laplace angle between Rosalind, Belinda and Perdita  is a sum of
the 43:44 resonant angle with Belinda/Perdita and the 7:8 resonant angle between Rosalind/Perdita,
so it is a resonant chain but involving a mean-motion resonance with the outer pair Rosalind/Perdita.
Rosalind, Belinda and Perdita  are near a low-integer 2:-3:1 
Laplace resonance and  Rosalind, Cupid and Perdita are near a 4:-7:5 Laplace resonance, 
and these could be strong
because many of their multiples would contribute to the resonance. 
The 5:-61:56 with Portia, Cupid and Belinda is a chain with the 5:6 between Portia and Cupid and
the 55:56 resonance with Cupid and Belinda.
Likewise the 4:-61:57 with Juliet, Cupid and Belinda is a resonant chain
(the 4:5 with Juliet/Cupid and the 56:57 with Cupid/Belinda).
Cupid and Belinda are so near each other that the 55:56 resonance is nearby 
even though the nearest resonance is the 57:58.
The three-body resonances involving Juliet and Portia perhaps account for the sensitivity of
Cupid's crossing timescale to the presence of bodies other than Belinda and Perdita \citep{french12}.

In Figures \ref{fig:combo_d} and \ref{fig:combo_c} we found histograms of Laplace angles exhibiting structure, and
they are resonant chains, but instead of involving mean motions between consecutive pairs,
they involve a mean-motion resonance between the inner and outer body of the trio.
There are two ways to create the three-body $p:-(p+q):q$ Laplace angle from a difference of first-order resonance arguments
involving pairs of bodies,
\begin{multline}
\theta = (p+q-1) \lambda_i -(p+q) \lambda_j + \varpi_i \\
- \left[ (q-1) \lambda_i - q \lambda_k + \varpi_i \right]  \label{eqn:theta_bracket_left}
\end{multline} 
for the $(p+q-1):(p+q)$ resonances between bodies $i,j$ and the $(q-1):q$ resonance between
bodies $i,k$ and 
\begin{multline}
\theta = p \lambda_i -(p+1) \lambda_k + \varpi_k \\
- \left[ (p+q) \lambda_j - (p+q+1) \lambda_k + \varpi_k \right]   \label{eqn:theta_bracket_right}
\end{multline} 
for the $p:(p+1)$ resonance between bodies $i,k$ and the $(p+q):(p+q+1)$ resonance between
bodies $j,k$.
The 20:-27:7 resonance with Desdemona, Juliet and Rosalind is an example of that in equation \ref{eqn:theta_bracket_left}
and the 7:-43:36 Laplace angle between Rosalind, Belinda and Perdita is an example of that in equation \ref{eqn:theta_bracket_right}.

\section{Three-body resonant strengths and chaotic behavior near a resonant chain of two first-order mean-motion resonances}
\label{sec:calc}

From the Laplace angle histograms,
we have identified candidate three-body resonances in the Uranian system.
While many of the variations in orbital elements in the Cupid, Belinda and 
Perdita trio appear to be caused by a trio of two-body resonances,
three-body resonances seem particularly important amongst the Bianca, Cressida, Desdemona, Juliet and Portia group.
In section \ref{sec:3per} we calculate, using a near-identity canonical transformation,
three-body resonance strengths for the setting where
a trio of bodies is near (but not extremely close to) a pair of two-body first-order mean-motion resonances.
Three-body resonance strengths and their libration frequencies are computed for the strong three-body resonances 
previously identified in the Uranian satellite system.

When a two-body resonant angle freezes, this gives a small divisor in the near-identity canonical transformation
used in section \ref{sec:3per},
so in section \ref{sec:twotwo} we employ a different canonical transformation for a Hamiltonian containing
two first-order resonant terms.  The resulting Hamiltonian resembles a forced pendulum and is used to estimate Lyapunov timescales
from resonant overlap in the setting when a trio of bodies is in a resonant chain of two first-order resonances.
 
\subsection{Resonant strengths of three-body resonances near two-body first-order mean-motion resonances}
\label{sec:3per}

\citet{quillen11}  ignored the effect of nearby two-body resonances
when estimating the strength of a three-body resonance. 
However, Figures \ref{fig:CresJulPor} and \ref{fig:CresDesPor} suggest that 
these are stronger than three-body resonances
that are distant from two-body resonances.
To estimate the strength of resonant-chain three-body resonances 
we follow a similar procedure to that used by \citet{quillen11}, using
a first-order (in perturbation strengths) near-identity canonical transformation.  However,
instead of using zeroth-order perturbation terms (in eccentricity)
we use first-order (in eccentricity) perturbation terms.
Here we consider the case when the system is near, but not in, either two-body resonance so
that small divisors do not invalidate the first-order nature of the transformation.  

We consider the Keplerian Hamiltonian, precession terms due to the oblate planet 
and two first-order (in eccentricity) resonance terms
\begin{multline}
H ( \vec\Lambda, \vec \Gamma, \vec \lambda, \vec \gamma) = 
\sum_l \left[ - \frac{m_l^3}{2 \Lambda_l^2} +  B_l \Gamma_l \right] \\
 \epsilon_p \Gamma_j^{\frac{1}{2}} \cos (p \lambda_j + (1-p)\lambda_i - \varpi_j)\\
+ \epsilon_q \Gamma_j^{\frac{1}{2}} \cos (q \lambda_k + (1-q)\lambda_j - \varpi_j) \label{eqn:H22}
\end{multline}
with 
\begin{eqnarray}
\epsilon_p(\Lambda_i,\Lambda_j) &=& - \frac{m_i m_j^3}{\Lambda_j^2} \left( \frac{2}{\Lambda_j} \right)^\frac{1}{2}
f_{31} ( \alpha_{ij},p)  \nonumber \\
&=& 
- \frac{m_i m_j^\frac{1}{2} 2^\frac{1}{2}}{a_j^\frac{5}{4} }
f_{31} ( \alpha_{ij},p) \nonumber \\
\epsilon_q (\Lambda_j, \Lambda_k) &=&  - \frac{m_j m_k^3}{\Lambda_k^2} \left( \frac{2}{\Lambda_j} \right)^\frac{1}{2}
f_{27} ( \alpha_{jk},q) \nonumber \\
&=&
- \frac{m_j^ \frac{1}{2} m_k 2^\frac{1}{2}}{a_k a_j^\frac{1}{4} }
f_{27} ( \alpha_{ij},p) \label{eqn:epsilonpq}
\end{eqnarray}
using equations \ref{eqn:Vij} and \ref{eqn:epsilonij} for the coefficients for the two-body first-order mean-motion resonances.
We define angles
\begin{eqnarray}
\phi_p  &\equiv & p \lambda_j + (1-p)\lambda_i - \varpi_j \nonumber \\
\phi_q  & \equiv & q \lambda_k + (1-q)\lambda_j - \varpi_j  .
\end{eqnarray}
We have chosen two resonant angles that contain $\varpi_j$.
The Hamiltonian contains two terms that are first order in perturbation parameters $\epsilon_p, \epsilon_q$.

Using a canonical transformation first order in perturbation strengths, we try to remove the two resonant terms.
The result is a Hamiltonian that contains no first-order terms but does contain second-order 
terms proportional to $\epsilon_p \epsilon_q$.
We use  
a generating function that is a function of new momenta ($\vec \Lambda', \vec \Gamma'$) and old angles
($\vec \lambda, \vec \gamma$) 
\begin{multline}
F_2(\vec \Lambda' , \vec \Gamma'; \vec \lambda, \vec \gamma) = 
\sum_l \left[  \Lambda_l'  \lambda_l + \Gamma_l' \gamma_l \right] \\
- \frac{\epsilon_p  \Gamma{'}_j^\frac{1}{2} }{\dot \phi_p }  \sin  \phi_p
-   \frac{\epsilon_q  \Gamma{'}_j^\frac{1}{2} }{\dot \phi_q }  \sin  \phi_q
\label{eqn:F2p}
\end{multline}
with divisors
\begin{eqnarray}
\dot \phi_p &\equiv &p n_j + (1-p) n_i + B_j \nonumber \\
\dot \phi_q &\equiv &q n_k + (1-q) n_j + B_j 
\end{eqnarray}
and with $B_j$ from secular perturbations.  The 
mean motions, $B_j$, $\epsilon_p$ and $\epsilon_q$ are evaluated using momenta $\vec \Lambda'$  .
Near a two-body resonance $\dot \phi_p$ or $\dot \phi_q$ is small, leading to a strong
perturbation or a small divisor.
We assume here that the system is near but not exactly on resonance so these divisors
never actually reach zero.  Equivalently we assume that the angles $\phi_p, \phi_q$ are circulating, increase
or decrease continually,
and do not librate around a particular value or remain fixed.
In the next section we will employ a different change of variables that contains no small divisors.

The canonical transformation gives a near-identity transformation.  New coordinates are equivalent
to old coordinates plus a term that is first order in perturbation strengths $\epsilon_p$ or $\epsilon_q$.
Relations between new and old coordinates are
\begin{eqnarray}
\Lambda_i &=& \frac{\partial F_2}{\partial \lambda_i} 
  = \Lambda_i' -  (1-p) \frac{\epsilon_p  }{\dot \phi_p }  \Gamma{'}_j^\frac{1}{2} \cos \phi_p \nonumber \\
\Lambda_j &=& \frac{\partial F_2}{\partial \lambda_j}
 = \Lambda_j' 
         -  p       \frac{\epsilon_p }{\dot \phi_p } \Gamma{'}_j^\frac{1}{2}\cos \phi_p 
         - (1-q)  \frac{\epsilon_q }{\dot \phi_q } \Gamma{'}_j^\frac{1}{2}\cos \phi_q   \nonumber \\
\Lambda_k &=& \frac{\partial F_2}{\partial \lambda_k} 
 =   \Lambda_k' -  q \frac{\epsilon_q  }{\dot \phi_q }  \Gamma{'}_j^\frac{1}{2} \cos \phi_q \nonumber \\
\lambda_i' &=& \frac{\partial F_2}{\partial \Lambda_i'} 
= \lambda_i +  
  \left [  \frac{\epsilon_p }{\dot \phi_p } \frac{ \partial  \dot \phi_p }{\partial \Lambda_i} - 
  \frac{ \partial  \epsilon_p }{\partial \Lambda_i}  \right]
  \frac{ \Gamma{'}_j^\frac{1}{2} }{\dot \phi_p } \sin \phi_p \nonumber \\
\lambda_j' &=& \frac{\partial F_2}{\partial \Lambda_j'}
 = \lambda_j +  
      \left [  \frac{\epsilon_p }{\dot \phi_p } \frac{ \partial  \dot \phi_p }{\partial \Lambda_j} - 
      \frac{ \partial  \epsilon_p }{\partial \Lambda_j}  \right]
      \frac{ \Gamma{'}_j^\frac{1}{2} }{\dot \phi_p} \sin \phi_p \nonumber \\
      && +  
      \left [  \frac{\epsilon_q }{\dot \phi_q } \frac{ \partial  \dot \phi_q }{\partial \Lambda_j} - 
      \frac{ \partial  \epsilon_q }{\partial \Lambda_j}  \right]
      \frac{ \Gamma{'}_j^\frac{1}{2} }{\dot \phi_q} \sin \phi_q \nonumber \\
\lambda_k' &=& \frac{\partial F_2}{\partial \Lambda_k'}
 = \lambda_k +   
      \left [  \frac{\epsilon_q }{\dot \phi_q } \frac{ \partial  \dot \phi_q }{\partial \Lambda_k} - 
      \frac{ \partial  \epsilon_q }{\partial \Lambda_k}  \right]
      \frac{ \Gamma{'}_j^\frac{1}{2} }{\dot \phi_q} \sin \phi_q \nonumber  .
\end{eqnarray}
\begin{eqnarray}
\gamma_i' &=& -\varpi_i' = \frac{\partial F_2}{\partial \Gamma{'}_i} = \gamma_i = - \varpi_i \nonumber \\
\gamma_j' &=& -\varpi_j' = \frac{\partial F_2}{\partial \Gamma{'}_j} 
= - \varpi_j -   \frac{ \epsilon_p } { 2 \Gamma{'}_j^\frac{1}{2} \dot \phi_p }\sin \phi_p 
-   \frac{ \epsilon_q } { 2 \Gamma{'}_j^\frac{1}{2} \dot \phi_q }\sin \phi_q \nonumber \\
\gamma_i' &=&  \frac{\partial F_2}{\partial \Gamma{'}_k} = \gamma_k = - \varpi_k \nonumber 
\end{eqnarray}
\begin{eqnarray}
\Gamma_i &=& \frac{\partial F_2} {\partial \gamma_i} = \Gamma_i' \nonumber \\
\Gamma_j &=& \frac{\partial F_2} {\partial \gamma_j} 
= \Gamma_j' - \frac{ \epsilon_p } {\dot \phi_p} \Gamma{'}_j^\frac{1}{2} \cos \phi_p
                    - \frac{ \epsilon_p } {\dot \phi_q} \Gamma{'}_j^\frac{1}{2} \cos \phi_q \nonumber \\
 \Gamma_k &=& \frac{\partial F_2} {\partial \gamma_k} = \Gamma_k' \nonumber \\
 \label{eqn:f2p_coord}
\end{eqnarray}

Inserting the new variables into the Hamiltonian (equation \ref{eqn:H22}) we expand
to second order in perturbation strengths $\epsilon_p$ and $\epsilon_q$. 
We neglect terms proportional to $\cos^2 \phi_p$ or $\sin^2 \phi_p$ (and similarly for $\phi_q$) 
and keep   terms proportional to $\cos \phi_p \cos \phi_q$
and $\sin \phi_p \sin \phi_q$.  We rewrite these products in terms of the Laplace angle 
\begin{equation}
\theta \equiv \phi_q - \phi_p = (p-1)\lambda_i - (p-1 + q) \lambda_j + q \lambda_k 
\end{equation}
that is similar to that discussed in the previous section  where we discussed a search for nearby
three-body resonances
 (see equation \ref{eqn:theta} but with $p-1$ replacing $p$).

Neglecting the primes on the coordinates, the Hamiltonian  (equation \ref{eqn:H22}) in the new variables  is 
\begin{multline}
K ( \vec\Lambda, \vec \Gamma, \vec \lambda, \vec \gamma) = 
\sum_l \left[ - \frac{m_l^3}{2 \Lambda_l^2} +  B_l \Gamma_l \right] \\
+ \chi_{pq} \cos ((p-1)\lambda_i - (p+q-1) \lambda_j + q \lambda_k).
\label{eqn:K22}
\end{multline}
The first-order terms (proportional to  $\epsilon_p$ or $\epsilon_q$) have been
removed leaving a single three-body term that is second order in perturbation strengths and proportional to $\epsilon_p \epsilon_q$.
The three-body term has coefficient
\begin{multline}
\chi_{pq} = 
- \frac{3}{2} \frac{p(1-q) \Gamma_j}{m_j a_j^2} \frac{\epsilon_p \epsilon_q}{\dot \phi_p \dot \phi_q} \\
+(q-1) \frac{ \epsilon_q\Gamma_j }{2 \dot \phi_p  }
     \left(
     \frac{ \epsilon_p}{\dot \phi_p} \frac{\partial \dot \phi_p }{\partial \Lambda_j} 
    - \frac{\partial \epsilon_p} {\partial \Lambda_j}
    \right) \\ 
- p \frac{ \epsilon_p \Gamma_j }{2 \dot \phi_p }
    \left(
    \frac{ \epsilon_q}{\dot \phi_q} \frac{\partial \dot \phi_q }{\partial \Lambda_j} 
    - \frac{\partial \epsilon_q} {\partial \Lambda_j}
     \right) \\ 
- \frac{p \epsilon_p \Gamma_j }{2 \dot \phi_p} \frac{\partial \epsilon_q}{\partial \Lambda_j} 
+ \frac{(q-1) \epsilon_q \Gamma_j }{2 \dot \phi_q} \frac{\partial \epsilon_p}{\partial \Lambda_j} \\
-  \frac{\epsilon_q \epsilon_p}{2 }  \left( \frac{1}{\dot \phi_p} + \frac{1}{\dot\phi_q}\right).
\label{eqn:chi_pq_a}
\end{multline}
The first term arises from the Keplerian part of the Hamiltonian, the remainder from the resonant terms.
The second and third terms come through perturbations on mean longitudes,
the fourth and fifth terms through perturbations on $\Lambda$, and the last term
from perturbations on $\varpi_j$ and $\Gamma_j$.
Neglecting the dependence of precession rates on $\Lambda_j$,
\begin{eqnarray}
\frac{\partial \dot \phi_p}{\partial \Lambda_j} &\approx& (p-1) \frac{3 n_j}{\Lambda_j} = \frac{3(p-1)}{m_j a_j^2} \\
\frac{\partial \dot \phi_q}{\partial \Lambda_j} &\approx & -q \frac{3 n_j}{\Lambda_j} = -\frac{3q}{m_j a_j^2} 
\end{eqnarray}
and we use this
 to simplify $\chi_{pq}$ to  
\begin{multline}
\chi_{pq} \approx 
 \frac{9pq}{2}\frac{\epsilon_p \epsilon_q}{\dot \phi_p \dot \phi_q} \frac{ \Gamma_j}{m_j a_j^2}  
 -  \frac{\epsilon_q \epsilon_p}{2 }  \left( \frac{1}{\dot \phi_p} + \frac{1}{\dot\phi_q}\right).\label{eqn:chi_pq}
\end{multline}
The last term in equation \ref{eqn:chi_pq}, independent of $\Gamma_j$, dominates because it does not depend on 
the square of the eccentricity of the $j$-th body.
This term only arises if both of the two first-order resonant terms are proportional $\Gamma_j^\frac{1}{2}$.
If we had chosen first-order resonances with arguments
$p \lambda_j +  (1-p) \lambda_i  - \varpi_i$ and $q \lambda_k +  (1-q) \lambda_i  - \varpi_j$,
the estimated three-body resonance strength would not have contained a term independent of
eccentricity.

In the low eccentricity and low mass setting,
\citet{quillen11} suspected that first-order resonance terms could be neglected when estimating
a three-body resonance strength, precisely due to their expected dependence on eccentricity.
The first term  in equation \ref{eqn:chi_pq} does depend on eccentricity so the 
eccentricity-independence of the
 last term is unexpected.

We try to understand why one of the terms in equation \ref{eqn:chi_pq} is independent of momentum $\Gamma_j$
by the considering an `indirect' effect (see section 4 by \citealt{nesvorny98}).  For example,
\citet{nesvorny98} considered  
the perturbations on the asteroidÕs motion that are raised by the oscillations of JupiterÕs orbit forced by Saturn.
Recall the  Hamiltonian in equation \ref{eqn:K4d}. 
We focus on only the term associated with the $p$ resonance or $\epsilon_p \Gamma_j^\frac{1}{2} \cos \phi_p$.
Hamilton's equation (neglecting the $q$ resonance) gives
\begin{equation}
\dot \Gamma_j = -\frac{\partial H}{\partial \gamma_i} = \epsilon_p \Gamma_j^\frac{1}{2} \sin \phi_p
\end{equation}
that we rewrite as 
\begin{equation}
\frac{d}{dt}  \Gamma_j^\frac{1}{2} = \frac{\epsilon_p }{2} \sin \phi_p.
\label{eqn:dtGam}
\end{equation}
If the angle $\phi_p$ circulates, we can integrate this to give
\begin{equation}
\Gamma_j^\frac{1}{2} = \frac{\epsilon_p}{2\dot \phi_p} \cos \phi_p  + {\rm constant} .
\end{equation}
When inserted into the other resonant term, $ \epsilon_q \Gamma_j^\frac{1}{2} \cos \phi_q$, 
we gain a three-body term
\begin{multline}
\frac{ \epsilon_q \epsilon_p}{2\dot \phi_p} \cos \phi_p \cos \phi_q  = 
 \frac{ \epsilon_q \epsilon_p}{4\dot \phi_p} \left[ \cos (\phi_p - \phi_q)  +  \cos (\phi_p + \phi_q) \right] .
\end{multline}
The three-body term is independent of eccentricity or $\Gamma_j$.
Here we essentially followed the estimates for three-body resonance strengths
in the asteroid belt by \citet{murray98}, where the presence of Saturn introduces 
additional frequencies into Jupiter's orbit and these give the three-body resonances.

Using equation \ref{eqn:epsilonpq}, and neglecting terms proportional to $\Gamma_j$, we can write equation \ref{eqn:chi_pq} 
for the three-body resonance strength as
\begin{eqnarray}
\chi_{pq} \sim -\frac{ m_i m_j m_k}{a_j^\frac{3}{2} a_k}  f_{31}(\alpha_{ij},p) f_{27}(\alpha_{jk},q) 
\left( \frac{1}{\dot \phi_p} + \frac{1}{\dot \phi_q} \right) 
\end{eqnarray}
and using equation \ref{eqn:fapprox} for $f_{27}$ and $f_{31}$ for closely-spaced bodies
\begin{eqnarray}
\chi_{pq} \sim \frac{ m_i m_j m_k}{a_j^\frac{3}{2} a_k} \frac{1}{16 \delta_{ij} \delta_{jk} }
e^{- p \delta_{ij} - q \delta_{jk}}
\left( \frac{1}{\dot \phi_p} + \frac{1}{\dot \phi_q} \right).
\label{eqn:chi_approx}
\end{eqnarray}

To estimate the strength of the three-body resonance we use the same canonical transformation as  in section 3.2
of \citet{quillen11}.
The generating function
\begin{multline}
F_2 (\vec \lambda, \vec J) = J((p-1) \lambda_i - (p-1+q)\lambda_j + q \lambda_k)\\ + \lambda_j J_j + \lambda_k J_k
\label{eqn:F2J}
\end{multline} 
gives in vicinity of resonance
\begin{equation}
H(J,\theta) = \frac{A_\theta J^2}{2} + b_\theta J + \chi_{pq} \cos \theta + ...
\end{equation}
Here the new momentum 
\begin{equation}
J  = \frac{\Lambda_i}{p-1} \label{eqn:Jsize}
\end{equation}
 is conjugate to the Laplace angle $\theta$.
The coefficients are
\begin{eqnarray}
A_\theta &=& -3 \left( \frac{(p-1)^2}{m_i a_i^2} + \frac{(p-1+q)^2}{m_j a_j^2} + \frac{ q^2}{m_k a_k^2}  \right)\\
b_\theta &=& (p-1)n_i - (p-1+q) n_j + q n_k  \label{eqn:ABtheta}
\end{eqnarray}
(using equations 32 and 33 of \citealt{quillen11}).
The frequency $b_\theta$ describes distance to resonance.
This frequency can be recognized as equivalent to $\dot \theta$ that we used earlier
 (equation \ref{eqn:dottheta} but with index $p-1$ replacing index $p$).

The three-body resonant libration frequency $\nu_3$  can be estimated from $\chi_{pq}$ and $A_\theta$
\begin{equation}
\nu_3 \approx \sqrt{|A_\theta \chi_{pq}| } \label{eqn:nu_3}
\end{equation}
and the condition to be in the vicinity of resonance is 
\begin{equation}
\left| \frac{b_\theta}{\nu_3} \right| \lesssim 1
\end{equation}
following section 3.3 of \citet{quillen11}.
From the resonance separatrix width, we estimate the
 size of  variations of momentum 
\begin{equation}
\Delta J \sim 2\sqrt{ \frac{\chi_{pq}}{A_\theta}} =  \frac{2\nu_3}{A_\theta} \label{eqn:DeltaJ}
\end{equation}
and using equation \ref{eqn:Jsize}, related variations in semi-major axis of the $i$-th body
\begin{equation}
\delta_i \sim  \frac{2 (p-1)}{m_i a_i^{1/2}} \Delta J. \label{eqn:delta_i}
\end{equation}
with $\delta_i \equiv \Delta a_i/a_i$.
Conserved quantities  
\begin{eqnarray}
(p-1)J_j &=& (p-1) \Lambda_j + (p+q) \Lambda_i \nonumber \\
(p-1) J_k &=& (p-1) \Lambda_k - q \Lambda_i 
\end{eqnarray}
 relate motions between the semi-major axes of
consecutive bodies with the outer and inner two bodies moving together and the middle one moving
in the opposite direction, or 
\begin{eqnarray}
- m_j \delta_j = m_i \frac{(p+q-1)}{p-1} \delta_j = m_k \frac{(p+q-1)}{q}  \delta_j. \label{eqn:deltaijk}
 \end{eqnarray}

\citet{quillen11} estimated three-body resonance strengths, $\epsilon_{pq}$, from two zeroth-order (in eccentricity) terms.
Here we estimate three-body resonance strengths, $\chi_{pq}$, from two first-order terms.
We can compare the computed resonance strengths by comparing $\epsilon_{pq}$ to $\chi_{pq}$.
Equations 23 and 46 of \citet{quillen11} give
\begin{equation}
\epsilon_{pq} \sim \frac{m_i m_j m_k}{12 \delta_{ij} \delta_{jk}} \ln \delta_{ij} \ln \delta_{jk} 
\exp(-(p \delta_{ij} + q \delta_{jk})).
\end{equation}
Taking a ratio of equation \ref{eqn:chi_approx} to this we estimate
\begin{equation}
\frac{\chi_{pq}}{\epsilon_{pq}} \sim \frac{1}{\ln \delta_{ij} \ln \delta_{jk}} 
\left( \frac{1}{\dot \phi_p} + \frac{1}{\dot \phi_q} \right).
\end{equation}
For a system near a two-body resonance, the divisor $\dot \phi_p$ or $\dot \phi_q$ would
have a larger magnitude than the logarithmic terms.
Consequently we expect a three-body resonance that is a resonant chain comprised of nearby two-body resonances 
to be stronger than that comprised of two single zeroth-order terms.  

For a $p:-(p+q):q$ resonance comprised of zeroth-order terms the $2p:-2(p+q):2q$ resonance
has approximately the same size coefficient
as long as $p\delta_{ij} \lesssim 1$ and $q \delta_{jk} \lesssim 1$; 
in other words $\epsilon_{pq} \sim \epsilon_{2p,2q} \sim \epsilon_{up,uq}$ for integer $u$.
This is not true when combining first-order resonances, $\chi_{pq} \ne \chi_{2p,2q}$.
For example, if the system is
near the $p:p+1$ first-order resonance,  then it is not near the first-order $2p:2p+1$ resonance,
but it would be near the second-order $2p:2p+2$ resonance. 
Combining two second-order (in eccentricity) resonant
terms, and following the same procedure to remove the first-order (in perturbation strength) terms via
canonical transformation, 
produces a resonant term that would depend on eccentricity ($\propto \Gamma_j$). We expect
three-body resonance strengths estimated from two second-order resonance terms
 to be weaker than those estimated from pairs of first-order resonances.

We use $\chi_{pq}$ to estimate the frequencies associated with the resonant
chain three-body resonances identified in our numerical integration.
Listed in Table \ref{tab:chain} are the minimum distances to three-body resonance, $\dot \theta$ or $b_\theta$ (see equations, \ref{eqn:dottheta} or \ref{eqn:ABtheta}),
during the integration for $t<10^{12}$s.  At the time of minimum distance to resonance,
we computed $\nu_3$ using equation \ref{eqn:nu_3}, based on the resonant strength
from first-order terms, $\chi_{pq}$ (equation \ref{eqn:chi_pq}).  
We also computed the three-body libration frequency using $\epsilon_{pq}$ (based on zeroth-order terms
and computed using equation 23 of \citealt{quillen11}).  
The ratio of the two frequencies is also listed and shows that the resonance strengths computed
using first-order terms can be an order of magnitude higher for the resonant chain three-body resonances
than previously computed using zeroth-order terms alone.

Table \ref{tab:chain} shows that the libration frequencies of the strongest three-body resonances are at most one to two orders
of magnitude smaller than the frequencies of the two-body resonances.  The strongest three-body resonance, the
46:-59:13 with Cressida, Desdemona and Portia, 
has an estimated libration period of only 3 years, and this is only  a few times longer than the
libration period in the Cressida/Desdemona 46:47 mean-motion resonance (see Table \ref{tab:twobodystuff}).
The three-body resonances are surprisingly strong considering that they must be second order in perturbation
strengths and perturbation strengths are  weak because of the low masses of the inner Uranian moons.
This resonance strength is because of the
 small inter-body separations, the small divisors, $\dot \phi_q$ or $\dot \phi_q$, and the lack of dependence
on eccentricity.  Checking the distance to resonance
we find that the minimum distance to resonance, $|\dot \theta|$, is in some cases
less than the resonance libration frequency, implying that there are times during the integration when
the three-body resonances are important.

The canonical transformation we used (equation \ref{eqn:f2p_coord}) contains small divisors $\dot \phi_p, \dot \phi_q$.
At what point is the near-identity canonical transformation no longer a good approximation?
The $p$ resonance  has a characteristic frequency scale given in equation \ref{eqn:omij}.  
Taking as a limit the smallest possible $\dot \phi_p$ to be equal to $\nu_p$ (the characteristic frequency
associated with the $p$ resonance) and inserting this value into the eccentricity independent term for the three-body 
resonance strength
 (equation \ref{eqn:chi_pq})
we estimate
\begin{eqnarray}
\chi_{pq} &\sim& \frac{\epsilon_p \epsilon_q}{\dot \phi_p}
\sim \frac{\epsilon_p \epsilon_q}{\nu_p}  
\sim \frac{\epsilon_p \Gamma_j^\frac{1}{2}\epsilon_q \Gamma_j^\frac{1}{2}}{\nu_p \Gamma_j} \nonumber \\
&\sim& \epsilon_q \Gamma_j^\frac{1}{2}
\end{eqnarray}
and we have used equation \ref{eqn:varepsilon} for the characteristic energy scale of the $p$ resonance.
This is equal to the energy in the $q$ resonance.   As long as  $|\dot \phi_p|$ and $|\dot \phi_q|$
are smaller than the respective $p$ or $q$ resonance libration frequency, the system
is not in the vicinity of the $p$ or $q$ resonance, and the canonical transformation is valid.
Just outside this region we estimate that the three-body resonance strength approaches
that of the two-body resonances and  the three-body resonance
strengths can be nearly as strong as the two-body resonance strengths (and consistent with our calculated values).

\begin{table*}
\begin{minipage}{166mm}
\vbox to 170mm{\vfil
\caption{\large Three-body  resonant chains   \label{tab:chain} }
\begin{tabular}{@{}lll crrr rrrrrrrr}
\hline
(1)     &   (2)              &    (3)         & (4)             &  (5)                        &   (6)                         &(7)       &(8)    & (9)    & (10) & (11) &(12) &(13)&(14) \\
$i$     &    $j$            & $k$          & q:-(p+q):q  & $\dot\theta_{init}$ & $\dot \theta_{m} $ & $\nu_3$(Hz)  &$\nu_{r}$  
& $\lambda_{olp}$  & $\lambda_{olp,m}$ & $\mu_\epsilon$ & $\delta_i$ & $\delta_j$ & $\delta_k$ \\
\hline
   Bianca &  Cressida & Desdemo. & 15:-62:47 & 8.4e-08 & -1.7e-10 &  2.7e-08 &   14.0 &     0.8 &   -0.002 &  -0.16 & 1.5e-06 & 1.8e-06 & 2.1e-06  \\ 
   Bianca & Desdemo. &    Juliet & 11:-36:25 & 6.7e-07 & 5.8e-07 &  5.3e-09 &    6.0 &     8.9 &    7.675 &  -0.07 & 6.4e-07 & 9.4e-07 & 1.8e-07  \\ 
 Cressida & Desdemo. &    Portia & 46:-59:13 & -2.6e-07 & 3.6e-11 &  7.6e-08 &   10.2 &    -2.4 &    0.000 &  -0.78 & 3.9e-06 & 8.0e-06 & 2.1e-07  \\ 
 Cressida &    Juliet &    Portia & 15:-39:24 & 1.9e-06 & 1.8e-06 &  9.6e-09 &    3.3 &    16.3 &   15.271 &  -0.14 & 1.2e-06 & 1.5e-06 & 3.9e-07  \\ 
Desdemo. &    Juliet &  Rosalind & 20:-27:7 & 3.4e-08 & 1.2e-10 &  1.0e-09 &    1.2 &     0.5 &    0.002 &  -0.32 & 2.8e-07 & 1.1e-07 & 8.6e-08  \\ 
 Cressida & Desdemo. &    Juliet & 45:-70:25 & -9.5e-07 & -4.9e-07 &  3.1e-08 &    5.5 &    -8.9 &   -4.610 &  -0.89 & 1.2e-06 & 3.0e-06 & 3.0e-07  \\ 
Desdemo. &    Juliet &    Portia & 23:-47:24 & 2.2e-07 & 2.6e-08 &  1.2e-08 &    3.3 &     1.9 &    0.221 &  -0.13 & 2.0e-06 & 1.2e-06 & 2.5e-07  \\ 
Desdemo. &    Juliet &  Rosalind & 23:-31: 8 & 9.0e-07 & 7.4e-07 &  2.0e-09 &    2.3 &    12.4 &   10.178 &  -0.33 & 4.9e-07 & 1.9e-07 & 1.5e-07  \\ 
   Portia &  Rosalind &   Belinda & 13:-24:11 & -1.7e-07 & -1.7e-07 &  1.5e-09 &    2.4 &    -3.9 &   -3.931 &  -0.18 & 4.3e-08 & 5.7e-07 & 1.3e-07  \\ 
   Portia &     Cupid &   Belinda &  5:-61:56 & 1.2e-06 & 3.9e-07 &  1.3e-08 &    1.0 &     5.7 &    1.841 &  -0.42 & 4.5e-10 & 2.4e-06 & 1.8e-08  \\ 
   Juliet &     Cupid &   Belinda &  4:-61:57 & -7.3e-07 & -5.7e-07 &  9.8e-09 &    1.0 &    -3.4 &   -2.647 &  -0.15 & 6.2e-10 & 1.8e-06 & 1.3e-08  \\ 
 Rosalind &   Belinda &   Perdita &  8:-49:41 & 8.1e-08 & 9.7e-11 &  2.4e-09 &    1.4 &     0.4 &    0.001 &  -0.21 & 6.4e-09 & 1.9e-08 & 6.6e-07  \\ 
 Rosalind &   Belinda &   Perdita &  9:-55:46 & 4.2e-07 & 5.8e-08 &  3.2e-09 &    1.8 &     2.0 &    0.270 &  -0.20 & 7.5e-09 & 2.2e-08 & 7.7e-07  \\ 
 Rosalind &     Cupid &   Belinda & 11:-73:62 & 8.8e-07 & 1.4e-09 &  5.2e-09 &    1.4 &     3.8 &    0.006 &  -0.10 & 1.9e-09 & 8.0e-07 & 5.4e-09  \\ 
 Rosalind &     Cupid &   Belinda & 10:-67:57 & -4.9e-07 & -3.2e-07 &  8.9e-09 &    2.4 &    -2.3 &   -1.486 &  -0.10 & 3.6e-09 & 1.5e-06 & 1.0e-08  \\ 
 Rosalind &     Cupid &   Perdita & 11:-38:27 & 5.9e-07 & 3.6e-08 &  3.0e-10 &    2.6 &    35.1 &    2.139 &  -0.10 & 3.5e-10 & 7.6e-08 & 1.8e-08  \\ 
    Cupid &   Belinda &   Perdita & 57:-101:44 & -5.5e-07 & 4.1e-10 &  5.9e-08 &   92.3 &    -2.5 &    0.002 &  -0.45 & 9.5e-06 & 1.3e-07 & 2.4e-06  \\ 
   Juliet &    Portia &  Rosalind & 22:-33:11 & 9.4e-07 & 9.4e-07 &  3.4e-09 &    2.1 &     8.2 &    8.177 &  -0.14 & 5.4e-07 & 3.5e-07 & 8.3e-07  \\ 
   Juliet &    Portia &  Rosalind & 44:-66:22 & 1.9e-06 & 1.9e-06 &  3.7e-09 &    2.4 &    10.1 &   10.216 &  -0.14 & 3.0e-07 & 1.9e-07 & 4.5e-07  \\ 
\hline
Desdemo. &    Juliet &  Rosalind & 20:-27: 7 & 3.4e-08 & 1.2e-10 &  1.4e-09 &    1.7 &     0.4 &    0.001 &   0.07 & 3.9e-07 & 1.5e-07 & 1.2e-07  \\ 
 Rosalind &     Cupid &   Perdita &  7:-24:17 & 1.2e-06 & 8.8e-07 &  9.5e-09 &   89.0 &    78.3 &   56.142 &   0.05 & 1.8e-08 & 3.9e-06 & 8.9e-07  \\ 
 Rosalind &   Belinda &   Perdita &  7:-43:36 & -2.6e-07 & -4.0e-09 &  7.6e-09 &    4.4 &    -1.3 &   -0.019 &   0.10 & 2.3e-08 & 6.9e-08 & 2.4e-06  \\ 
\hline
\end{tabular} 
{\\ 
Columns 1-3.  The three bodies considered.
Col 4.  A three-body angle $\theta = p \lambda_i - (p+q)\lambda_j + \lambda_k$  is defined with integers p,-(p+q),q. 
Except for the bottom three rows,
the chain consists of bodies $i,j$ in a first-order p:p+1 mean-motion resonance and the bodies $j,k$ in a q-1:q 
mean-motion resonance. In the bottom three rows the chain arises from  p+q-1:p+q with bodies $i,j$ and q-1:q for bodies 
$i,k$ (Desdemona, Juliet and Rosalind  20:-27:7)
or p+q:p+q+1 for bodies $j,k$ and p:p+1 for bodies $i,k$ 
(Rosalind, Cupid and Perdita  7:-24:17 and Rosalind, Belinda and Perdita 7:-43:36), 
(see equations \ref{eqn:theta_bracket_left} and \ref{eqn:theta_bracket_right}).
Col 5.  Distance to three-body resonance, $\dot \theta$, at the start of the numerical integration in Hz.
Col 6.  Minimum distance to three-body resonance, $\dot \theta$, for $t<10^{12}$s in Hz.
Col 7.  Libration frequency in Hz of the three-body resonance.
Here $\nu_3$ refers to the libration frequency  computed with $\chi_{pq}$, 
using equations \ref{eqn:nu_3}, \ref{eqn:chi_pq} and with $A_\theta$ from equation \ref{eqn:ABtheta}.
Col 8. The ratio of the libration frequency computed with $\chi_{pq}$ compared to that computed with $\epsilon_{pq}$.
Col 9.  Overlap ratio for the two first-order resonances computed from the initial $\dot \theta_{init}$.
The libration frequency of the stronger resonance  is used to compute this ratio.
Col 10.  Overlap ratio computed from minimum $\dot \theta_{min}$.
Col 11. Ratio  of resonance strengths for the two first-order resonances.  Listed is
$\epsilon_q/\epsilon_p$ if the $p$ resonance is stronger
otherwise $\epsilon_q/\epsilon_p$ is given.  
Col 12-15.  Sizes of variations in semi-major axis for each body (equations \ref{eqn:delta_i}, \ref{eqn:deltaijk})
in units of semi-major axis.
\\
}}  
\end{minipage}
\end{table*}

\subsection{Distance to resonant chains}

For three satellites with inter-body spacings $\delta_{ij}$ and $\delta_{jk}$, what are the properties of
the nearest resonant chain?
The closest first-order resonance to the pair of bodies $i,j$ and to the pair of bodies $j,k$ have 
integers $p,q$ such that
\begin{eqnarray}
p &\sim& \frac{2}{3} \delta_{ij}^{-1} \nonumber \\
q &\sim &\frac{2}{3} \delta_{jk}^{-1} 
\end{eqnarray}
(using equation \ref{eqn:nij}).
What is the frequency of the resonant angle for the $p+1 $  first-order resonance?  
The difference between the two frequencies $|\dot \phi_p  - \dot \phi_{p+1} | = n_{ij} \sim \frac{3}{2}\delta_{ij}$.  
This allows us to estimate the maximum possible value of $|\dot \phi_p|$ for the closest first-order resonance.
We find
\begin{eqnarray}
|\dot \phi_p | < \frac{2}{3} \delta_{ij} \nonumber\\
|\dot \phi_q | < \frac{2}{3} \delta_{jk}
\end{eqnarray}
for the nearest first-order resonances.
Subtracting the two frequencies, $\dot \phi_p - \dot \phi_q$,  
we find that the frequency of the associated Laplace angle satisfies
\begin{equation}
|\dot \theta| < \frac{2}{3} \left( \delta_{ij} + \delta_{jk}\right).
\end{equation}

These values of integers $p,q$ give a slowly moving Laplace angle, but they may not give the slowest Laplace angle.
However if $\delta_{ij} < \delta_{jk}$ then we can increase or decrease the $p$ index to find a slower three-body angle
and vice-versa for the $q$ index.  For example, in our integration, Cupid and Belinda are pretty near
the 56:57 resonance even though the closest first-order resonance is the 57:58.
There might be other integers $p,q$ giving very small  values for $|p n_i - (p+q) n_j + qn_k|$ but 
these might not be near the $p:p+1$ and $q:q+1$ resonances.
If we wanted the slowest Laplace angle we could use 
Dirichlet's approximation theorem 
to estimate
a maximum value of $|\dot \theta|$ for the closest three-body resonance.
We estimate that
there is a pair of integers $p,q$ with the first pair of bodies near the $p$ first-order resonance
and the second pair near the $q$ first-order resonance, minimizing $\dot \theta$, with 
\begin{equation}
|\dot \theta| <  \min(\delta_{ij} ,\delta_{jk}),
\end{equation}
$|\dot \phi_p |  \lesssim \delta_{ij} $ and $|\dot \phi_q | \lesssim \delta_{jk}$.
For this choice of $p,q$
\begin{equation}
\left| \frac{1}{\dot \phi_p} + \frac{1} {\dot \phi_q} \right| \gtrsim \max ( \delta_{ij}^{-1} , \delta_{jk}^{-1} ) 
\end{equation}
giving for the three-body resonance strength
\begin{equation}
|\chi_{pq}| \gtrsim \frac{m_i m_j m_k}{ \delta_{ij} \delta_{jk}   \min ( \delta_{ij} , \delta_{jk})} 
\end{equation}
using equation \ref{eqn:chi_approx}. 
Let us call $m_1$ the most massive of our three masses, $m_2$ the middle one and $m_3$ 
the least massive.   Let $\delta_1$ be the smaller of $\delta_{ij}$ and $\delta_{jk}$ and $\delta_2$ 
the larger one. Using this notation
\begin{equation}
|\dot \theta| \lesssim \delta_1. \label{eqn:dt_1}
\end{equation}
The coefficient (equation \ref{eqn:A_theta}) is inversely proportional to the mass of the lightest body and contains the square of
$p^2$ or $p^2$ or $(p+q)^2$ depending upon which one is associated with the lowest mass body.
Conservatively $|A_\theta| \gtrsim  \delta_2^{-2} m_3^{-1}$.
The   three-body resonance libration frequency (equation \ref{eqn:nu_3})
\begin{equation}
\nu_3 \gtrsim \sqrt{m_1 m_2} \delta_1^{-1} \delta_2^{- 3/2 }. \label{eqn:nu3_approx}
\end{equation} 
Using equation \ref{eqn:DeltaJ}
we can estimate a characteristic scale for semi-major axis variations in 
resonance for the lightest body
\begin{equation}
da/a \gtrsim \sqrt{m_1 m_2} \delta_1^{-1} \delta_2^{-1/2}. \label{eqn:daa}
\end{equation}
Conserved quantities can be used to estimate variations in semi-major axes for other bodies.
For masses similar to a few times $10^{-9}$ and separations of order 0.02, we estimate semi-major axis variations
(as a fractional change)
have a scale in the range $10^{-6}-10^{-7}$.  One of the divisors $\dot \phi_p$ 
or $\dot \phi_q$ could be smaller, giving a larger value for $\chi_{pq}$ and $da/a$.
This size scale is consistent with the size of small variations in semi-major axis seen during the integration
(see Figure \ref{fig:geom_ae}) and the sizes for semi-major axis variations in three-body resonant chains
 listed in Table \ref{tab:chain}.
The small variations in semi-major axis seen in the simulation might be attributed
to a continual state of diffusion in semi-major axes via weak, but ubiquitous, three-body resonances.
However the large variations in orbital elements cannot be attributed to the three-body resonances alone.

A crude diffusion coefficient for wander in semi-major axis due to the three-body resonant chains
can be estimated from the sizes of semi-major axis variations and the resonant libration frequency. 
The product of the square of equation \ref{eqn:daa} with equation \ref{eqn:nu3_approx}
gives a diffusion coefficient (due to three-body resonances) for variations in semi-major axis
\begin{equation}
D_a \sim m^3 \delta^{-11/2}
\end{equation}
and we have used a single mass and separation for the estimate.
This estimate has the same exponent for mass as that estimated by \citet{quillen11} (her equation 65) but
has a larger negative exponent for $\delta$ as the terms used to construct the three-body arguments are first- rather than 
zeroth order in  eccentricity. 
 If the wander in semi-major axes is due to first-order two-body resonances alone
then using similar scaling we would expect $D_a \propto m^2$ and so a weaker dependence on mass than estimated here.
The crossing timescale measured by \citet{french12} has the exponent for mass ranging from   -2 to -5.
If the system must diffuse an equivalent distance in all simulations before two moons cross orbits then the crossing timescale
would be proportional to the inverse of the diffusion coefficient.    The inverse of our estimated diffusion coefficients would
match only the shallowest end of the measured exponents.
Perhaps the shallower exponent measured for the Cupid/Belinda crossing time ($\sim -3$) compared to the 
Cressida/Desdemona pair ($\sim -4$, see Table 5 of \citealt{french12})
can be attributed to a more important role for two-body resonances in that pair.

For moon masses
 higher than those used in the integration studied here, the two-body resonances would be even more important as the system
 would be more likely to be found in a two-body resonance.
 We guess that for higher moon masses crossing timescales would be less strongly dependent on moon mass (and have a shallower
 exponent) but this is opposite to what was found (see Figure 5 of \citealt{french12}).
 As the moons wander in semi-major axis, the strengths of the three-body resonances
vary with proximity to first-order resonances.  The size of variations in orbital
elements and the time between variations could depend on proximity to first-order resonances.
In such a setting diffusion could be anomalous rather than ordinary. 
We find that simple diffusion estimates based on three-body resonances fail to 
 predict the numerically observed crossing times or trends seen in them.

\subsection{Resonance overlap for two first-order mean-motion resonances between two pairs of bodies}
\label{sec:twotwo}

The canonical transformation (equations \ref{eqn:F2p}, \ref{eqn:f2p_coord}) contains  divisors
$\dot \phi_p$ and $\dot \phi_q$ that could be small.
The transformation is no longer a near-identity transformation when the angles $\dot \phi_p$ and $\dot \phi_q$
do not circulate.
We consider again the Hamiltonian  in equation \ref{eqn:K22}, containing two first-order resonant terms for two pairs of bodies,
but now use a different canonical transformation
lacking any small divisors.
We first expand near-constant values of mean motion
$\vec \Lambda = \vec \Lambda_0 + \vec y$
where $\vec \Lambda_0$ gives mean motions $a_{i0}, a_{j0}, a_{k0}$.
The Hamiltonian (equation \ref{eqn:K22}) becomes
\begin{multline}
K(\vec y, \vec \Gamma; \vec \lambda, \vec \gamma) = 
\sum_{l=i,j,k} \left[ n_l y_l- \frac{3 y_l^2 }{2 m_l a_{l0}^2} + B_l \Gamma_l \right] 
\\
+ \epsilon_p \Gamma_j^\frac{1}{2} \cos (p \lambda_j  + (1-p) \lambda_k - \varpi_j)\\
+ \epsilon_q \Gamma_j^\frac{1}{2} \cos (q \lambda_k  + (1-q) \lambda_j - \varpi_j), \label{eqn:K22_exp}
\end{multline}
where the mean motions, $n_l$, correspond to those associated with $a_{i0}, a_{j0}, a_{k0}$.
We use a generating function that is a function of old angles and new momenta
\begin{multline}
F_2(J_i, J_j, J,\Gamma_j', \lambda_i, \lambda_j, \lambda_k, \gamma_j ) =\\
\Gamma_j' ( p \lambda_j + (1-p) \lambda_i - \varpi_j)  \\
+ J ( (p-1) \lambda_i  - (p-1+q)  \lambda_j  + q \lambda_l)  \\+ 
J_i \lambda_i + J_j \lambda_j \label{eqn:F2both}
\end{multline}
The canonical transformation gives relations between new and old coordinates 
\begin{eqnarray}
y_i &=&(\Gamma_j' - J)(1-p)  + J_i  \nonumber \\
 y_j&=&\Gamma_j'p - (p-1+q) J + J_j   \nonumber\\
y_k&=&Jq    \nonumber\\
\phi_p &=& p \lambda_j + (1-p) \lambda_i - \varpi_j  \nonumber\\
\theta &=&  (p-1) \lambda_i  - (p-1+q)  \lambda_j  + q \lambda_k  \nonumber\\
\lambda_i' &=& \lambda_i \nonumber \\
\lambda_j' &=& \lambda_j \nonumber\\
\Gamma_j' &=& \Gamma_j.  \label{eqn:F2both_coord}
\end{eqnarray}
Here the angle $\phi_p$ is conjugate to the momentum $\Gamma_j$ and 
the Laplace angle, $\theta$, is conjugate to the momentum $J$.  

The Hamiltonian (equation \ref{eqn:K22_exp}) in the new coordinates (equation \ref{eqn:F2both_coord}) is
\begin{multline}
K(\Gamma_j, J, J_i, J_j; \phi_p, \theta, \lambda_i, \lambda_j)  = \\
\frac{A\Gamma_j^2}{2} + \frac{A_\theta J^2}{2} +  b_j \Gamma_j + b_\theta J  
+ c \Gamma_j J  \\
+ \epsilon_p \Gamma_j^\frac{1}{2} \cos \phi_p + \epsilon_q \Gamma_j^\frac{1}{2} \cos (\theta + \phi_p)
\label{eqn:Kboth}
\end{multline}
and we have neglected $\Gamma_i, \Gamma_k$ as
they are conserved in this restricted setting.  We have dropped the primes from $\Gamma_j$ and the mean longitudes as they are  not changed by the transformation.
Coefficients are
\begin{eqnarray}
A_\theta &=& - 3 \left[ \frac{(p-1)^2}{m_i a_{i0}^2} + \frac{(p-1+q)^2}{m_j a_{j0}^2}  + \frac{q^2}{m_k a_{k0}^2} \right] \nonumber \\
b_\theta &=& n_i (p-1) - n_j(p-1+q) + n_k q \nonumber \\
c &=& \frac{3 (p-1)^2}{m_i a_{i0}^2} +  \frac{3 p(p-1+q)}{m_j a_{j0}^2} \label{eqn:A_theta}
\end{eqnarray}
and $A$ and $b_j$ are given in equation \ref{eqn:AB_2}.
Additional constants that depend on the conserved quantities $J_i, J_j$ have been dropped
as they can be removed by shifting $a_{i0}, a_{j0}, a_{k0}$.
Here $b_\theta$ gives proximity to the three-body resonance, and $b_j$ gives proximity 
to the $p-1:p$ first-order resonance between bodies $i,j$, as discussed in section \ref{sec:twobody}.

Manipulating equation \ref{eqn:F2both_coord} 
\begin{eqnarray}
J_i &=& y_i + (p-1) \Gamma_j - \frac{(p-1) y_k}{q} \nonumber \\
J_j &=& y_j - p\Gamma_j  + \frac{(p-1+q)y_k}{q}. 
\end{eqnarray}
As the new Hamiltonian (equation \ref{eqn:Kboth}) is independent of $\lambda_i,\lambda_j$,
the momenta $J_i,J_j$, are conserved quantities.
These conserved quantities relate motions of semi-major axes and the eccentricity of the middle
body.     The first conserved quantity implies that  motions of the inner and
outer body and the eccentricity of the middle body are coupled.  
Adding together 
\begin{equation}
J_i + J_j = y_i + y_j + y_k - \Gamma_j
\end{equation}
implying that all three bodies can move outwards if the middle body decreases in eccentricity. 


As  
in section \ref{sec:resover}, we attempt to approximate the Hamiltonian (equation \ref{eqn:Kboth})
so that it resembles the well-studied periodically-forced non-linear pendulum.
Assuming a mean value for $\Gamma_j$ we approximate an energy $\varepsilon_p = \epsilon_p \Gamma_j^\frac{1}{2}$
and define an energy ratio $\mu = \epsilon_q/\epsilon_p$.  We assume that $\dot \theta$ is never zero and that $J< \Gamma_j$.
This allows us to neglect the terms proportional to $J \Gamma_j$ and $J^2$.
The resulting Hamiltonian  is 
\begin{multline}
K(\Gamma_j, J, J_i, J_j; \phi_p, \theta, \lambda_i, \lambda_j)  = 
\frac{A\Gamma_j^2}{2}  +  b_j \Gamma_j + b_\theta J   \\
+ \varepsilon_p \left[\cos \phi_p + \mu \cos (\theta + \phi_p) \right]
\label{eqn:Kboth_short}
\end{multline}
and $\dot \theta = b_\theta$.
This Hamiltonian  resembles the well-studied periodically-forced non-linear pendulum (e.g., \citealt{chirikov79,shev02})
and suggests that
the separatrix of the $p$ resonance can be chaotic due to forcing by the $q$ resonance term.
The two resonances are separated by the frequency $b_\theta$.   Recall that the parameter $b_\theta$ we used previously 
to describe distance to a resonant-chain three-body resonance. Here it sets the distance between
the $p$ and $q$ resonances, each between a different pair of bodies. As $\dot \theta$ is the
difference between the frequencies of the $p$ and $q$ resonant angles, it
serves as a perturbation frequency in analogy to 
the forced pendulum model.

Lyapunov timescales are estimated in terms of a unitless
overlap ratio, $\lambda_{olp}$, that is the ratio between the  perturbation frequency
 and the frequency of small oscillations
of the dominant  first-order mean-motion resonance.
  Here 
\begin{equation}
\lambda_{olp} = \frac{b_\theta}{\nu_p},
\end{equation}
where $\nu_{p}$ is a characteristic libration frequency typical of the $p$ resonance
(equation \ref{eqn:numax}).
As seen in the Hamiltonian (equation \ref{eqn:Kboth})
the relative resonance strengths are set by the ratio $\epsilon_q/\epsilon_p$.
While our canonical transformation (equation \ref{eqn:F2both}) was chosen for a dominant
$p$ resonance,   if the $q$ resonance is stronger then we would have chosen its angle to be a coordinate.
A similar canonical transformation would give an overlap parameter that depends on the
frequency of libration in the $q$ resonance, $\lambda_{lop} = \frac{b_\theta}{\nu_q}$, and
the relative resonance strength would be $\epsilon_p/\epsilon_q$.

For resonant chains (pairs of first-order mean-motion resonances) we list in Table \ref{tab:chain}
the distances to the three-body resonances (here serving as the perturbation frequency in the analogy to
the forced pendulum)  at the beginning
of the integration and the minimum value measured in the time interval $0< t<10^{12}$s.
We also list overlap ratios, $\lambda_{olp}$, computed from both values of $b_\theta$ 
 using the libration frequency of the stronger resonance (that with larger libration frequency).
The strength ratio is also listed for each pair of resonances.

In Table \ref{tab:chain} we see that the ratio of resonance strengths for many of the resonance pairs is of order 1,
so if the overlap parameter is in the vicinity of 1/2 the  resonances overlap.
Most of the resonant chains
are comprised of consecutive pairs: the $p:p+1$ resonance between bodies $i,j$
and $q-1:q$ resonance between bodies $j,k$.   However, we list similar values for a few chains
where the chain is comprised of a consecutive resonant pair  and the outer and inner body in resonance
(as shown in equations \ref{eqn:theta_bracket_left} and \ref{eqn:theta_bracket_right}).

Table \ref{tab:chain} shows that overlap ratios vary from large to small values.
When $\lambda_{olp} \gg 1$ the width of the separatrix and energy perturbation size scale
in the separatrix are exponentially truncated (e.g. see equations 7 and 8 of \citealt{shev02} and \citealt{chirikov79}).
In the adiabatic regime, the separatrix width shrinks as it depends on $\lambda_{olp}^2$ and the 
Lyapunov timescale, in units of the libration period, is inversely proportional to $\lambda_{olp}$ \citep{shev08}.
The Lyapunov timescale approaches the libration perturbation period in the intermediate regime $\lambda_{olp} \sim 1/2$.

Table \ref{tab:chain} shows that
the following resonant pairs are in the regime of $\lambda_{olp}$ ranging from a few to near zero:
\begin{itemize}
\item Bianca/Cressida 15:16 and Cressida/Desdemona 46:47
\item Cressida/Desdemona 46:47 and Desdemona/Portia 12:13
\item Desdemona/Juliet 23:24 and Juliet/Portia 23:24
\item Desdemona/Juliet 20:21 and Juliet/Rosalind 6:7
\item Rosalind/Belinda 8:9 and Belinda/Perdita 40:41
\item Rosalind/Belinda 9:10 and Belinda/Perdita 45:46
\item Rosalind/Cupid 11:12 and Cupid/Belinda 61:62
\item Cupid/Belinda 57:58 and Belinda/Perdita 43:44
\item Desdemona/Juliet 26:27 and Desdemona/Rosalind 6:7
\item Rosalind/Perdita 7:8 and Belinda/Perdita 43:44
\end{itemize}
The trios in this list are likely to be  in a regime where the Lyapunov timescale
is similar to the perturbation frequency, $b_\theta$.  The Lyapunov
timescale is only short when $\lambda_{olp} \sim 1/2$, and there it is 
of order the resonance libration period.  Consequently we expect that 
there are intervals when the Lyapunov timescale is of order 
 the resonance libration period, which as shown in Table \ref{tab:twobodystuff}
ranges from a year to 10 years.  The perturbation sizes caused by
the resonance coupling are a fraction of the energy of the resonances themselves
(as the ratios $\epsilon_q/\epsilon_p$ are in the range 0.1-1).   Consequently when the
system is in the $p$ resonance separatrix, we expect large and frequent energy perturbations.

If the Cressida/Desdemona or Cupid/Belinda pairs were at all times in the vicinity of a first-order
resonance separatrix then they would display large (of size 0.1 the first-order resonance energy) and frequent
(approximately 10 years for the Lyapunov timescale) perturbations in semi-major axis and eccentricity.
However, large variations in their orbital elements are seen only a few times during the 30,000 years 
shown in Figure \ref{fig:geom_ae}.  Either the intermittency is due to overlap of 
subterms (as discussed in section \ref{sec:resover}) or these pairs spend only a small fraction of
the integration in the vicinity of their separatrices.

\subsection{Low-index Laplace angles}
\label{sec:lowq}

We mentioned in section \ref{sec:search3} that we suspected that three-body resonances
with low indices could be strong because multiples of zeroth-order interaction terms
can contribute to their strength.
We can sum these multiple terms to improve upon our estimate for their strength.
For a low-integer slowly-moving Laplace angle, 
the Hamiltonian (equation \ref{eqn:H3})
\begin{multline}
H(\vec \Lambda,\vec \lambda) = -\sum_{l=1,2,3}  \frac{m_l^3}{2 \Lambda_l^2}  + \\
 \sum_{u=1}^{u_{max}} \epsilon_{up,uq} \cos (u (p \lambda_i - (p+q) \lambda_j + q \lambda_k )) 
\end{multline}
and we have included multiples of the Laplace angle.
After canonical transformation (using equation \ref{eqn:F2J})
\begin{multline}
H(J, \theta) = \frac{A_\theta}{2} J^2 + b_\theta J  +  \sum_{u=1}^{u_{max}} \epsilon_{up,uq} \cos (u \theta) .
\end{multline}
The resonance strengths $\epsilon_{up,uq}$ as estimated
from the zeroth-order interaction terms are independent of $u$
as long as $up\delta_{ij} <1$ and $uq\delta_{jk} <1$ (see equation 23 of \citealt{quillen11} and equation \ref{eqn:approx}).
The limiting $u_{max}$ is  the maximum value of integer $u$ for which these conditions are met.
Using only the lowest integer term, $u=1$, the frequency of small oscillations  is
\begin{equation}
\nu_{u=1} = \sqrt{A_\theta \epsilon_{pq}}.
\end{equation}
However, when all terms are included
\begin{eqnarray}
\nu_{umax} &=& \sqrt{A_\theta \epsilon_{pq}} \sqrt{\sum_{u=1}^{u_{max}} u^2} \nonumber \\
    & =&  \sqrt{A_\theta \epsilon_{pq}} \left[ \frac{ u_{max} (u_{max}-1)(2u_{max}-1) }{6} \right]^\frac{1}{2} \nonumber \\
    &  \sim&   \sqrt{A_\theta \epsilon_{pq}} u_{max}^\frac{3}{2} 3^{-\frac{1}{2}} \label{eqn:nu_umax}
\end{eqnarray}
using the formula for the sum of squares $\sum_{i=1}^n i^2 = n(n-1)(2n-1)/6$.

Libration frequencies computed using a sum of indices are listed in Table \ref{tab:lowq} for a series of low-index
Laplace resonances identified in our search for nearby three-body resonances (in section \ref{sec:search3}).
Table \ref{tab:lowq} contains distances to resonances, as computed for Table \ref{tab:chain}, along with
$u_{max}$, the largest integer that satisfies  $up\delta_{ij} <1$ and $uq\delta_{jk} <1$.
The table also lists the ratio of $\nu_{max}$ to $\nu_{u=1}$ that is computed solely from the lowest multiple.

We can see from Table \ref{tab:lowq} that there are times when trios of bodies are within the vicinity
of low-index Laplace resonances (e.g., the 3:-20:17 for Rosalind, Cupid and Belinda) and
that the libration frequencies are in some cases comparable to the fastest resonant chain
libration frequencies listed in Table \ref{tab:chain}.
Our suspicion that the low-index Laplace resonances could be comparatively strong (based on
structure present in their angle histograms) is supported.

\begin{table*}
\begin{minipage}{166mm}
\vbox to 170mm{\vfil
\caption{\large Low-index Laplace resonances   \label{tab:lowq} }
\begin{tabular}{@{}lll crrr rrrrr}
\hline
(1)     &   (2)              &    (3)         & (4)             &  (5)                        &   (6)                         &(7)                            &(8)    & (9)    &(10) &(11) &(12)       \\
$i$     &    $j$            & $k$          & q:-(p+q):q  & $\dot\theta_{init}$ & $\dot \theta_{min} $ & $\nu_{umax}$(Hz)  & $u_{max}$ & 
$\frac{\nu_{umax}}{\nu_{u=1}}$ & $\delta_i$ & $\delta_j$ & $\delta_k$\\
\hline
Cressida &    Juliet &    Portia &  5:-13:8 & 6.4e-07 & 6.0e-07 &  1.1e-08 &  5 &    5.5 & 4.2e-06 & 4.9e-06 & 1.3e-06  \\ 
   Bianca &  Cressida &    Juliet &  9:-19:10 & 3.4e-07 & 2.6e-07 &  1.4e-09 &  3 &    2.2 & 5.3e-07 & 3.2e-07 & 7.5e-08  \\ 
   Bianca & Desdemona &    Juliet &  7:-23:16 & -1.2e-07 & 2.6e-07 &  1.8e-09 &  3 &    2.2 & 3.4e-07 & 5.0e-07 & 9.9e-08  \\ 
 Cressida & Desdemona &    Juliet &  9:-14:5 & -1.9e-07 & -9.8e-08 &  3.4e-08 &  8 &   11.8 & 6.5e-06 & 1.6e-05 & 1.6e-06  \\ 
 Cressida & Desdemona &    Portia &  7:-9:2 & -3.0e-07 & -2.4e-07 &  6.1e-08 &  10 &   16.9 & 2.1e-05 & 4.2e-05 & 1.1e-06  \\ 
 Cressida & Desdemona &  Rosalind &  7:-8:1 & 4.9e-08 & 2.6e-08 &  2.3e-08 &  9 &   14.3 & 9.0e-06 & 1.6e-05 & 1.8e-06  \\ 
 Cressida &    Juliet &    Portia &  3:-8:5 & -7.7e-07 & -7.7e-07 &  1.7e-08 &  8 &   11.8 & 1.1e-05 & 1.3e-05 & 3.5e-06  \\ 
 Cressida &    Juliet &    Portia &  5:-13:8 & 6.4e-07 & 6.0e-07 &  1.1e-08 &  5 &    5.5 & 4.2e-06 & 4.9e-06 & 1.3e-06  \\ 
 Cressida &    Juliet &    Portia &  8:-21:13 & -1.3e-07 & -1.2e-07 &  5.5e-09 &  3 &    2.2 & 1.3e-06 & 1.6e-06 & 4.2e-07  \\ 
 Cressida &    Juliet &  Rosalind & 11:-17:6 & -3.9e-07 & -3.7e-07 &  4.3e-10 &  2 &    1.0 & 1.4e-07 & 9.8e-08 & 1.1e-07  \\ 
Desdemona &    Juliet &    Portia &  1:-2:1 & 2.6e-07 & 2.5e-07 &  7.4e-08 &  37 &  127.3 & 2.9e-04 & 1.6e-04 & 3.5e-05  \\ 
Desdemona &    Juliet &  Rosalind &  3:-4:1 & 8.7e-07 & 8.5e-07 &  8.4e-09 &  12 &   22.5 & 1.5e-05 & 5.9e-06 & 4.5e-06  \\ 
Desdemona &    Portia &  Rosalind &  1:-2:1 & 3.5e-07 & 3.4e-07 &  5.5e-09 &  18 &   42.2 & 2.2e-05 & 5.3e-06 & 1.9e-05  \\ 
   Juliet &    Portia &  Rosalind &  2:-3:1 & 8.5e-08 & 8.6e-08 &  2.1e-08 &  18 &   42.2 & 3.6e-05 & 2.3e-05 & 5.5e-05  \\ 
   Juliet &     Cupid &   Belinda &  1:-15:14 & 3.3e-07 & 1.3e-07 &  5.8e-08 &  7 &    9.5 & 1.5e-08 & 4.4e-05 & 3.2e-07  \\ 
   Portia &  Rosalind &   Belinda &  6:-11: 5 & 9.7e-07 & 9.6e-07 &  1.2e-09 &  3 &    2.2 & 7.3e-08 & 9.6e-07 & 2.2e-07  \\ 
   Portia &     Cupid &   Belinda &  1:-12:11 & 6.5e-07 & 4.9e-07 &  8.6e-08 &  8 &   11.8 & 1.5e-08 & 8.0e-05 & 5.9e-07  \\ 
 Rosalind &     Cupid &   Perdita &  2:-7:5 & -3.2e-07 & -2.9e-07 &  7.3e-10 &  8 &   11.8 & 4.6e-09 & 1.0e-06 & 2.4e-07  \\ 
 Rosalind &     Cupid &   Perdita &  6:-21:15 & -9.6e-07 & -8.6e-07 &  2.3e-10 &  3 &    2.2 & 5.0e-10 & 1.1e-07 & 2.5e-08  \\ 
 Rosalind &     Cupid &   Belinda &  1:-7:6 & -6.6e-07 & -6.4e-07 &  4.7e-08 &  15 &   31.9 & 1.7e-07 & 7.6e-05 & 5.2e-07  \\ 
 Rosalind &     Cupid &   Belinda &  2:-13:11 & 7.2e-07 & 5.4e-07 &  2.5e-08 &  8 &   11.8 & 5.2e-08 & 2.1e-05 & 1.4e-07  \\ 
 Rosalind &     Cupid &   Belinda &  3:-20:17 & 5.6e-08 & 2.8e-11 &  2.0e-08 &  6 &    7.4 & 2.7e-08 & 1.1e-05 & 7.5e-08  \\ 
 Rosalind &     Cupid &   Belinda &  4:-27:23 & -6.0e-07 & -5.3e-07 &  1.1e-08 &  4 &    3.7 & 1.1e-08 & 4.8e-06 & 3.2e-08  \\ 
 Rosalind &   Belinda &   Perdita &  1:-6:5 & 3.4e-07 & 3.0e-07 &  1.8e-08 &  13 &   25.5 & 3.9e-07 & 1.2e-06 & 4.0e-05  \\ 
 Rosalind &   Belinda &   Perdita &  7:-43:36 & -2.6e-07 & -4.0e-09 &  1.7e-09 &  2 &    1.0 & 5.1e-09 & 1.6e-08 & 5.4e-07  \\ 
    Cupid &   Belinda &   Perdita & 13:-23:10 & -3.3e-08 & 8.8e-10 &  3.3e-09 &  7 &    9.5 & 2.3e-06 & 3.3e-08 & 5.9e-07  \\ 
\hline
\end{tabular} 
{\\ 
Columns 1-3.  The three bodies considered.
Col 4.  A three-body angle $\theta = p \lambda_i - (p+q)\lambda_j + \lambda_k$  is defined with integers p,-(p+q),q. 
Col 5.  Distance to three-body resonance, $\dot \theta$, at the start of the numerical integration in Hz.
Col 6.  Minimum distance to three-body resonance, $\dot \theta$, for $t<10^{12}$s in Hz.
Col 7.  Libration frequency in Hz of the three-body resonance.
Here $\nu_{umax}$ refers to the libration frequency computed with equation \ref{eqn:nu_umax}.
Col 8. The maximum index $u_{max}$.
Col 9.  The ratio  of libration frequency computed from a sum of indices to that
only using the lowest one $\nu_{umax}/\nu_{u=1}$. 
Col 10-12. Sizes of variations in semi-major axis caused by the three-body resonance (equations \ref{eqn:delta_i}, \ref{eqn:deltaijk}). 
\\
}}  
\end{minipage}
\end{table*}

\section{Summary and Discussion}
\label{sec:sum}

By examining a numerical integration by \citet{french12}, 
we probed the resonant mechanisms responsible for the chaotic evolution 
of the inner moons in the Uranian satellite system.
We have identified strong first-order mean-motion resonances between
pairs of moons and estimated their characteristic libration frequencies using a perturbative
nearly-Keplerian Hamiltonian model for systems with multiple massive bodies.
Using histograms of slow-moving three-body resonant angles, we have found trios of bodies exhibiting
coupled motions when three-body angles freeze.
We find that histograms of three-body Laplace angles tend to show structure if
the angle is also a resonant chain (equal to the difference
between two first-order resonant angles between two pairs of moons).
Histograms of low-integer three-body Laplace angles also sometimes show structure.
The strongest three-body resonance identified is the 46:-57:13 between
Cressida, Desdemona and Portia, which is also near the 46:47 first-order mean-motion resonance
between Cressida and Desdemona and the 12:13 first-order mean-motion resonance between Desdemona
and Portia.  
Coupled motions between Cressida, Desdemona and Portia tend to 
take place when the three-body Laplace angle makes a transition from 0 to $\pi$ or vice versa.

Using a near-identity canonical transformation, we estimated the strength
of three-body resonances that are also resonant chains.  
We found that in some cases the three-body resonance 
libration frequencies are only one to two orders of magnitude smaller
than those of first-order resonances.  As gravitational interaction terms only involve two bodies,
three-body resonance strengths are second order in perturbation strength
(and so a higher power of moon mass).  Because they are sensitive to the separation between
bodies and the distance to a first-order resonance (serving as a small divisor)
and are independent of eccentricity, they can be nearly as strong as first-order
mean-motion resonances.  
We calculated that
low-integer three-body Laplace resonances can have similar-sized 
libration frequencies, with resonance strength due to the contribution of 
many multiples of the three-body angles arising from zeroth-order terms in the disturbing function.
For any trio of three closely-spaced bodies, we estimated the strength of the three-body resonance
associated with the nearest resonance chain.  We estimated
 associated semi-major axis variations and found them similar in size to
 the ubiquitous small variations seen in our simulation.  
This suggests that the small 
 coupled variations in semi-major axis,
seen throughout the simulation,
are due to ubiquitous and weak three-body resonant couplings. 

Using a canonical transformation without any small divisor, 
we considered the resonant chain 
setting where consecutive pairs of bodies are in two first-order resonances.
The transformed Hamiltonian resembles the well-studied forced pendulum model
but with the distance to three-body resonance (equivalently the time derivative 
of the Laplace angle)  serving as a perturbation frequency that
 describes an overlap between the two resonances.
We identified trios of bodies and associated
pairs of first-order resonances that are in a regime where short Lyapunov times (of order
a few times the resonance libration periods) are predicted.
When a pair of bodies is in the resonance separatrix, it can experience frequent (on the Lyapunov timescale) 
and large (approximately 0.1 the energy of the larger resonance) perturbations due to the resonance
between the other pair of bodies.
If the system spends long intervals in a resonance separatrix, then the system
could exhibit large jumps in orbital elements every libration period (or every few years for the Uranian satellites).   
However, the resonant angles associated with the first-order resonances for
Cupid/Belinda and Cressida/Desdemona instead exhibit
behavior that we might better describe as intermittent, experiencing large jumps in orbital elements
only a few times during the first 30,000 years of the simulation.
Subterms in each individual resonance are likely to be in an adiabatic regime and could account for
the intermittency.  Alternatively, if perturbations from a first-order resonance with a third body are responsible for the chaotic
behavior then perhaps the resonant pair spends only a small fraction of time in the vicinity of its separatrix
and this could account for the intermittency. 

\citet{quillen11} argued, based on the relatively small number of two-body
 resonances compared to three-body resonances (and this was also emphasized by \citealt{nesvorny98}), 
 that a closely-spaced multiple-planet system is unlikely to 
 be unstable due to two-body resonances alone.  
However, \citet{quillen11} estimated three-body resonance strengths using only zeroth-order terms 
and did not consider systems near or in two-body resonance.  
Here we find that the strongest three-body resonances are resonant chains and are near
  a pair of two-body resonances.  The higher
  strength of these resonances may alleviate some of the discrepancy between the predicted and
 numerically-measured three-body resonance strengths of \citet{quillen11}.
  
We found that
the strengths of three-body Laplace resonances associated with a resonant  chain are dependent on
small divisors.    
As the moons wander in semi-major axis, the strengths of these three-body resonances
vary with proximity to first-order resonances.   For the overlapping two-body resonances,
strong variations are likely only if one of the pairs of bodies is in the vicinity of its separatrix.
In such a setting, the size of variations in orbital
elements and the time between variations could depend on proximity to first-order resonances, and
diffusion can be anomalous (an associated random walk could be called  a L\'evy flight).
Although we have estimated the strengths of three-body resonances and Lyapunov timescales
for overlapping pairs of first-order resonances, we have tried but failed to account for power-law
relations measured for crossing timescales.   If the diffusion really is anomalous
then it will be challenging to develop a theoretical framework that can match the exponents measured
numerically for crossing times.

In this study we have neglected secular resonances as well as three-body resonances
that involve a longitude of pericenter of one of the bodies (such as
$12\lambda_{Des} - 49 \lambda_{Jul} + 38 \lambda_{Por} - \varpi_{Jul}$ that might
be related to the 49:51 second-order mean-motion resonance between Juliet and Portia).
We have also neglected the possibility that  
  a heavily-overlapped system (one with a near-zero overlap parameter), in 
  the 	adiabatic chaos regime described by \citet{shev08}
 (and so near a periodic orbit), might be integrable or
 stable \citep{lochak93} rather than chaotic.
 Chaotic behavior in this study has been crudely estimated via analogy to the
 periodically-forced pendulum.  However, exploration of Hamiltonian models
 containing only a few Fourier components could be used to better 
 understand the diffusive behavior.
 Despite our ability to estimate two- and three-body resonance strengths, we lack
 a mechanism accounting for
 the power law relations in numerically measured crossing timescales
in compact planar multiple-body systems.

We compare the role of three-body interactions in the inner Uranian satellite
to those in the asteroid belt (e.g., \citep{nesvorny98,nesvorny98b,murray98}).
In the asteroid belt, the distance (as a ratio in semi-major axis)
between bodies (asteroid and Saturn, asteroid and Jupiter, and Saturn and Jupiter)
is much larger  ($\Delta \sim 0.5$)  than the distances between
bodies in the inner Uranian satellite system ($\Delta \sim 0.02$).
The exponential decay of Laplace coefficients with
resonance index depends on the semi-major axis ratio of
two bodies. In the asteroid belt, the exponential decay
with resonance index makes the high index resonances weak.
In contrast in the inner Uranian satellite
the proximity of the bodies allows high index (but low order) resonances
to influence the dynamics.

Three-body resonance strengths are second order in planet mass.
Saturn and Jupiter have mass ratios that are approximately $10^5$ times larger
than the inner Uranian satellite mass ratios.
Because of higher body masses three-body resonances that are comprised
of second order terms in eccentricity can be strong in the asteroid belt.
This perhaps explains why three-body resonances
associated with Laplace angles that include a longitude of perihelion
can be important.   For example,
that associated with the angle $5\lambda_J -2\lambda_S -\lambda - \varpi$
with $\lambda_J$ and $\lambda_S$ the mean longitude of Jupiter and Saturn,
(see Figure 1 for asteroid 490 Veritas by \citealt{nesvorny98b}).
Because the masses and eccentricities are not low, chaos can arise from
overlap of multiplets in these resonances \citep{nesvorny98,murray98}.
We compare this to the Uranian satellite system where
the masses and eccentricities are so low that either
proximity to a first order resonance or a low index Laplace angle gives
a strong three-body resonance. In this setting overlap of first order
but high index mean motion resonances in pairs of bodies and weak three-body resonances
contribute to the chaotic behavior.


\vskip 0.3 truein

{\it Acknowledgements.}  
We thank Ivan Shevchenko, and Yanqin Wu for helpful correspondence and discussions.  
This work was supported in part by NASA under grant NNX13AI27G.
Robert S. French was supported by NASAÕs Planetary Geology and Geophysics Program under Grant NNX09AG14G. 
Additional support was provided by NASA through a Grant from the Space Telescope Science Institute, which is operated by the Association of Universities for Research in Astronomy, Incorporated, under NASA contract NAS5-26555.

\end{document}